\documentclass[reprint,10pt,twocolumn,footinbib,superscriptaddress,showkeys,prl,showpacs]{revtex4-1}
\usepackage{amsmath,amssymb}
\usepackage[dvips]{graphicx}
\usepackage[latin1]{inputenc}
\usepackage{color}
\usepackage{hyperref}
\hypersetup{backref,colorlinks=true,citecolor=blue}

\begin{document}
\title{Apparent Fracture in Polymeric Fluids under Step Shear}
\date{\today}
\author{Okpeafoh S. Agimelen}
\author{Peter D. Olmsted}
\affiliation{Soft Matter Physics Group, School of Physics and Astronomy, University of Leeds, Leeds, LS2 9JT, United Kingdom}

\begin{abstract}
  Recent step strain experiments in well-entangled polymeric liquids
  demonstrated a bulk fracture-like phenomenon.
We have studied  this
instability using a modern version of the Doi-Edwards theory for entangled polymers, and we find close quantitative agreement with the experiments. The phenomenon occurs because the viscoelastic liquid is sheared into a rubbery state that possesses an elastic constitutive instability (Marrucci and Grizzuti, 1983). The fracture is a transient manifestation of  this instability, which relies on the amplification of
  spatially inhomogeneous fluctuations. This mechanism differs from fracture in glassy materials and dense suspensions. 
  \end{abstract}



\maketitle
\textbf{\textit{Introduction--}} Viscoelastic liquids have slow timescales due to the relaxation of internal  degrees of freedom such as polymer deformation or the structures of self-assembled materials such as amphiphiles. These slow timescales give rise to dramatic effects, such as rubbery behaviour at high deformation rates and viscous behaviour at lower rates, and \textit{both}  solid-like or liquid-like features. Materials such as  amorphous solid polymers \cite{Doyle1972} or metallic glasses \cite{metallicglasses} have arguably the most dramatic behaviour possible for  a solid: rupture, fracture, and flow at a macroscopically sharp interface. This has been modelled as collective rupture of shear transformation zones (STZs) \cite{ManLan2007PSNSMP,*Manning2009rate}; and in  dense colloidal materials as due to the coupling between shear and density \cite{furukawa2009inhomogeneous}. 

Recent experiments have demonstrated  fracture-like behaviour in well-entangled polymeric \textit{liquids}.
Very rapid step strains were applied to polymer melts 
(\textit{e.g.} poly(styrene-butadiene) \cite{Boukany2009n2} or  poly(ethylene oxide) \cite{FangJOR2011}) with $Z \approx 53 - 160$
entanglements per polymer. At such high shear rates the liquid becomes rubbery and solid-like. After the step strain the solid-like melt relaxes homogeneously
for a short time, followed by a rapid relaxation during which the material splits into two layers moving in opposite
directions, separated by a thin ($\alt 40\,\mu\textrm{m}$)  shear band or `fracture'
layer [Fig.~1 of~\cite{Boukany2009n2}]. Ref.~\cite{Boukany2009n2} suggested that this is due to microscopic yield, such as a  sudden localized chain pull-out or loss of entanglements, perhaps analogous to the STZ picture for yield in amorphous solids \cite{furukawa2009inhomogeneous}. 

We show that these results can be explained by a pure constitutive instability due to the effects of shear flow on the elastic stress in the fluid, and is actually contained in the Doi-Edwards (DE) theory of entangled polymers \cite{doiedwards,Marrucci1983,*Morrison1992,Adams2009,*CaoLikhtmanPRL2012}; this  provides yet another mechanism for fracture, due purely to a constitutive shear instability in a viscoelastic \textit{liquid} brought suddenly into a (transient) solid state.

\begin{figure*}[tbh]
\centerline{\includegraphics[width=0.99\textwidth]{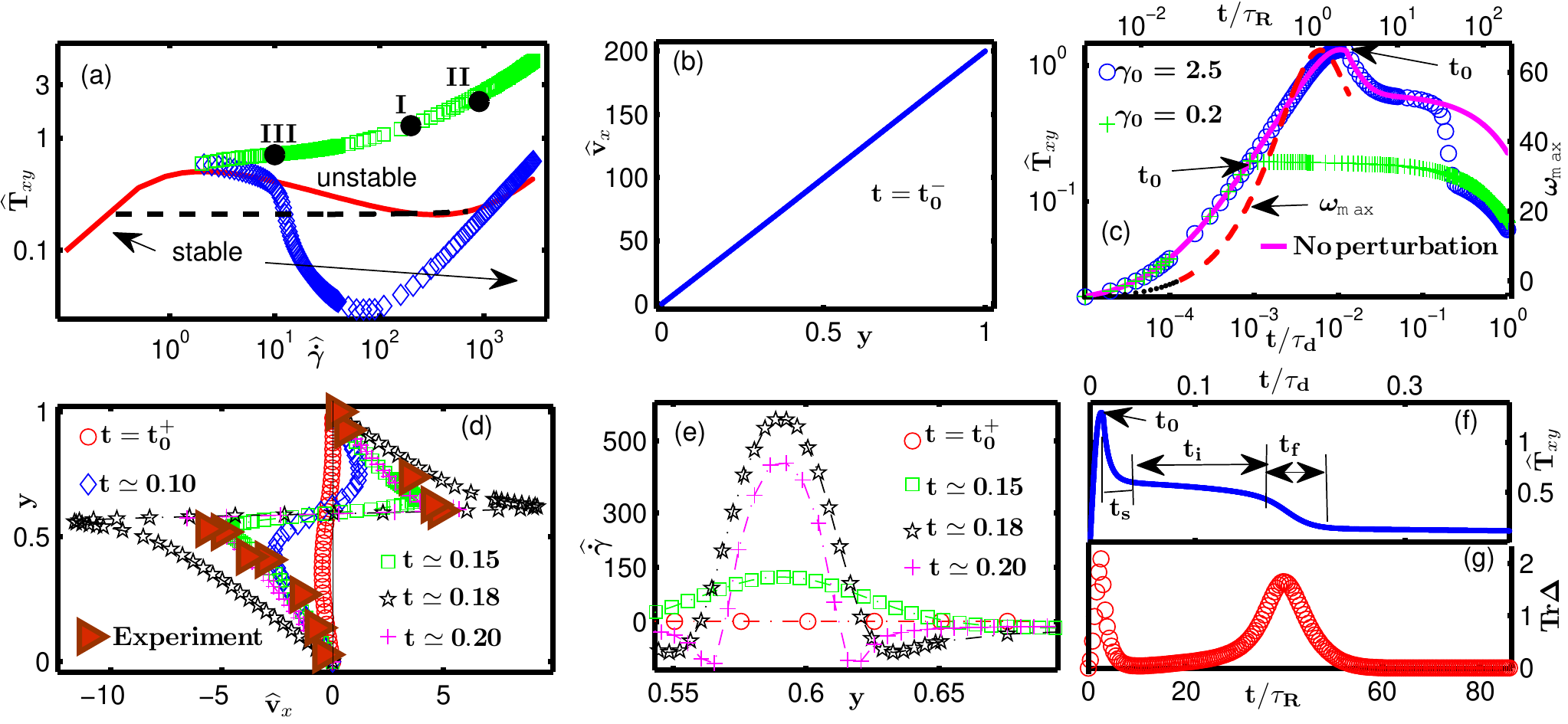}}
\caption{(a) Constitutive (solid line) and steady state shear banding
  (black dashes) curves. The stress overshoot is indicated by green
  squares $({\color{green} \square})$; small perturbations grow exponentially in time $(\omega_{\textrm{max}}>0)$ for stresses exceeding 
the   stress given by the blue diamonds $({\color{blue}\lozenge})$. The stresses at $t_0$ for three cases described in the text are indicated by I, II and III. (b)
  Velocity profile at $t_0^-$ (just before shear
  cessation) for $\langle\hat{\dot{\gamma}}\rangle=200$.  (c) Stress
  relaxation for step strains $\gamma_0=0.2, 2.5$; the solid line is
  for $\gamma_0=2.5$ with no initial perturbation.  The dot-dash line
  shows the evolution of the most unstable eigenvalue $\omega_{max}$,
  which becomes unstable $(\omega_{\textrm{max}}>0)$ in the red
  (dashed) region.  (d) Velocity profiles during fracture, with
  experimental data from~\cite{Boukany2009n2} superposed.  (e) Shear
  rate profiles, (f) stress relaxation, and (g) evolution of the
  maximum stretch in the gap $\textrm{Tr}\,\mathbf{\Delta}^{max}$.
  [Parameters: $Z = 72$, $\tau_R=\tau_d/216$,
  $\langle\hat{\dot{\gamma}}\rangle = 200$, $\gamma_0=2.5$,
  $t_0=0.01250\,\tau_d$, and $t_0^{\pm} = t_0 \pm 10^{-5}\tau_d$. Times $t$ and $1/\omega_{max}$ are displayed in units of $\tau_d$.]}
\label{fig1}
 \end{figure*}

The motion of an
entangled polymer is restricted to a tube-like region due to
the constraints imposed by surrounding chains. The DE theory for this
\cite{doiedwards} predicts a maximum in
the shear stress $T_{xy}$ as a function of shear rate
[Fig.~\ref{fig1}(a)], at a shear rate $\dot{\gamma}$ roughly equal to
the reciprocal of the time $\tau_d$ for a polymer to diffuse (or reptate) along its
tube. This non-monotonic constitutive behaviour  (which was not inferred in early experiments on polymer
melts~\cite{Menezes1982})
indicates instability, which can lead to inhomogeneous flows and \textit{shear
banding} \cite{Spenley1993,*Olmsted2008,*Lu2000}.  This constitutive
instability was widely implicated
\cite{huseby1966hypothesis,*lin1985explanation,*mcleish86,*mcleish87,*MNP91,*Denn90}
in the spurt effect \cite{Vinogradov1972}, responsible for
instabilities in industrial processes; however, spurt is now usually
attributed to wall slip
\cite{lim1989wall,*Wan1999PCE,*Den2001ARFM}. In rapid startup flow the
DE theory predicts the rubbery behaviour of a stress
overshoot
\cite{doiedwards,Marrucci1983,*Morrison1992,Adams2009,*CaoLikhtmanPRL2012}. Modern theories  incorporate chain
stretch and convected constraint release (CCR) -- chain relaxation due
to the release of entanglement constraints, which restores
stable constitutive behavior \cite{Ianniruberto1996,*Mead1998}.  
However, new observations of shear
banding  seem to validate the DE instability \cite{hu2007cre,*Ravindranath2008,*Tapadia2006n1,*Hu2010, Adams2009,Adams2009b,*Wang2009} in some cases. We will show that apparent `fracture' is another manifestation of the DE instability. 

 \begin{figure}[htb]
   \begin{center}
	\includegraphics[width=0.48\textwidth]{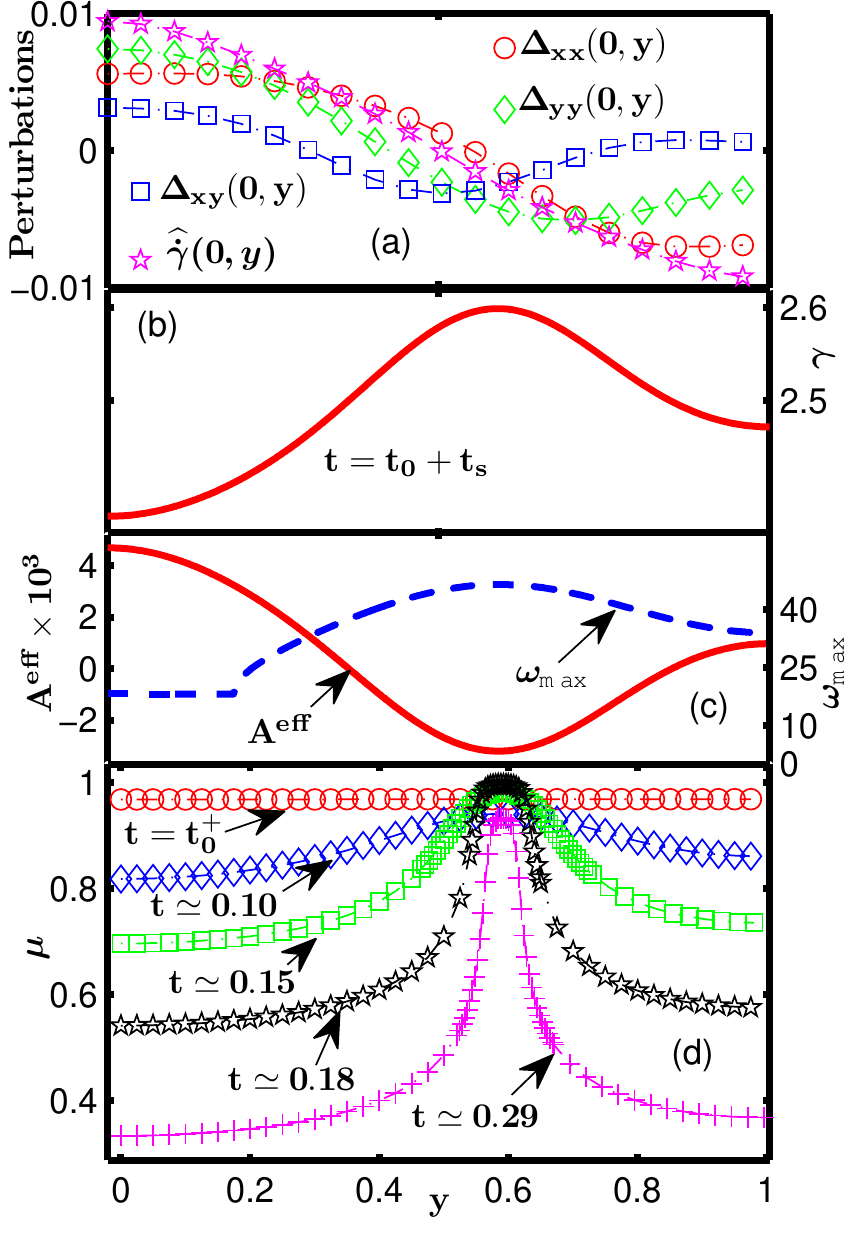}
   \end{center}
   \caption{Spatial profiles of (a) initial perturbation; (b) local strain;  (c) effective
     modulus ${\cal A\/}^{\textrm{eff}}$ as well as the unstable
     growth rate $\omega_{\textrm{max}}$, after cessation of flow and
     subsequent stretch relaxation; (d) Evolution of unrelaxed polymer
     segments $\mu(y,t)$ during fracture development. 
      [Parameters as in Fig.~\ref{fig1}. Time $t$ displayed
     in units of $\tau_d$.] 
   }
 \label{fig2}
 \end{figure}

\begin{figure*}[htb]
  \begin{center}
    \includegraphics[width=0.99\textwidth]{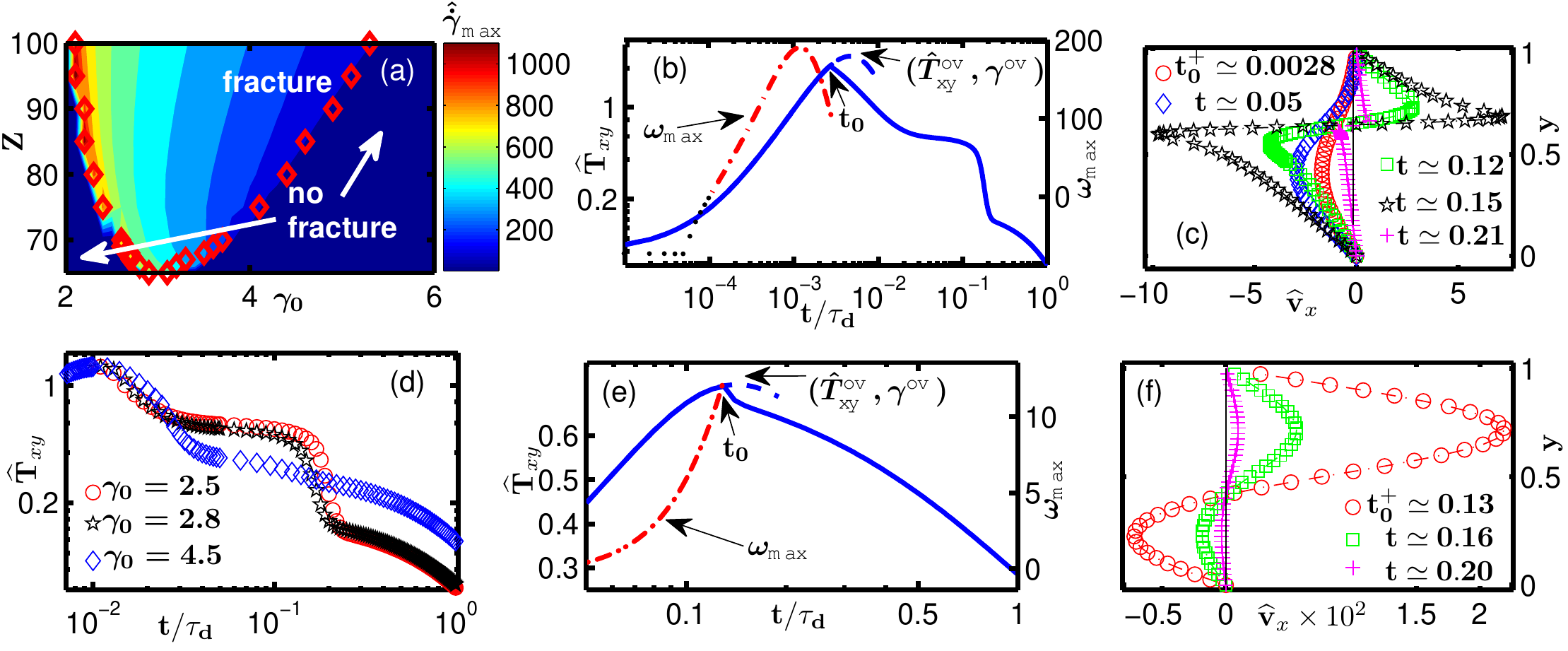}
  \end{center}
  \caption{(a)  Values of $Z$ and $\gamma_0$ required for
     fracture at fixed $\langle\widehat{\dot{\gamma}}\rangle$;
     contours show the maximum local shear rates during
     fracture. (bc) Case II ($\langle\hat{\dot{\gamma}}\rangle = 900,
    \langle\dot{\gamma}\rangle\tau_R = 4.2$, $\gamma_0=2.5$): (b)
    Stress relaxation and unstable growth rate
    $\omega_{\textrm{max}}$ (the dashed line is the stress overshoot with no initial perturbation); c) velocity profiles during
    fracture. (d) Stress decay for three different imposed strains
    $\gamma_0$. (ef) Case III ($\langle\hat{\dot{\gamma}}\rangle = 10,
    \langle\dot{\gamma}\rangle\tau_R = 0.046$, $\gamma_0=1.3$): (e)
    Stress relaxation and $\omega_{\textrm{max}}$, and (f) velocity
    profiles.  
    [All other parameters as in Fig.~\ref{fig1}. Time $t$ displayed in
    units of $\tau_d$.]  }
  \label{fig3}
\end{figure*}

\textbf{\textit{Model--}} We separate the total stress tensor
$\mathbf{T}$ into contributions from the polymer and a Newtonian
solvent, as $\mathbf{T} = G\mathbf{W} + \eta (\mbox{\boldmath{$\kappa$}} +
\mbox{\boldmath{$\kappa$}}^T) - p\mathbf{I},$
where $G$ is a modulus, $\eta$ is the solvent viscosity, the
pressure $p$ maintains incompressibility, $\mathbf{I}$ is the identity
tensor and $\kappa_{\alpha\beta} \equiv \partial v_{\alpha}/\partial
y_{\beta}$. The fluid velocity $\mathbf{v}$ (with no slip boundary conditions) with mass density $\rho$ obeys
\begin{equation}
  \rho\frac{d\mathbf{v}}{dt} \equiv \rho\left[\frac{\partial}{\partial t}  +
    \left(\mathbf{v}\cdot \nabla\right)\right]\mathbf{v} = \nabla
  \cdot\mathbf{T},
\label{eq3}
\end{equation}
where  $ \nabla\cdot\mathbf{T} = 0$ for
 very small Reynolds numbers, as is the case here.
The dimensionless polymeric conformation, or strain, tensor
$\mathbf{W}$ is assumed to obey the diffusive Rolie-Poly (RP)
model~\cite{Likhtman2003,Adams2009},
\begin{align}
\frac{d\mathbf{W}}{dt} & = \mbox{\boldmath{$\kappa$}} \cdot \mathbf{W}
+ \mathbf{W}\cdot\mbox{\boldmath{$\kappa$}}^T - \frac{1}{\tau
  _d}(\mathbf{W} - \mathbf{I}) - \frac{2\left(1
  -\sqrt{\frac{3}{\textrm{Tr}\mathbf{W}}}
 \right)}{\tau_R} \nonumber\\ 
& \quad \times \left(\mathbf{W} +
  \beta\sqrt{\frac{3}{\textrm{Tr}\mathbf{W}}}
  \left(\mathbf{W}  - \mathbf{I}\right)\right) + \mathcal{D}\nabla ^2 \mathbf{W}, \label{eq2}
\end{align}
which is a simplified form of the GLaMM mode, itself a modern version of Doi-Edwards theory 
\cite{Graham2003}. Here, $\tau _d$ is the reptation time, and the Rouse time $\tau
_R$ governs the relaxation of stretch $\textrm{Tr}(\mathbf{W})$. The
parameter $\beta$ quantifies CCR; a
large value of $\beta$ corresponds to more CCR, which leads to
monotonic (stable) behaviour of the shear stress. Spatial gradients due to stress `diffusivity' $\mathcal{D}$ are subject to the boundary condition $\nabla\mathbf{W} = 0$~\cite{Spenley1993,*Olmsted2008,*Lu2000}.  

 \textbf{\textit{Calculations--}} We consider two infinite flat plates
  separated by  $L\,\mathbf{\hat{y}}$ where the top plate  moves parallel to $\mathbf{\hat{x}}$ and the bottom
 plate is fixed. The velocity field is thus given by  $\mathbf{v} =
 v_x(t,y)\mathbf{\hat{x}}$, and $\mathbf{W} \equiv \mathbf{W}(t,y)$.  We define
 dimensionless quantities  $\hat{\dot{\gamma}} =
 \dot{\gamma}\tau _d$, $\widehat{\mathcal{D}} = \mathcal{D}\tau
 _d/L^2$, $\epsilon = \eta/(G\tau _d)$, $\hat{\rho} = \rho L^2/(G
 \tau _d ^2)$, $\hat{v} = \tau _d v /L$, and $\hat{t}=t/\tau_d$.  The
 degree of entanglement $Z$ determines the Rouse time via $\tau_R=\tau_d/(3Z)$ \cite{Likhtman2003,Graham2003}.  A desired
 average shear rate is imposed for a duration $t_0$ leading to a
 strain $\gamma _0 = \langle\hat{\dot{\gamma}}\rangle t_0$.

 The values $\tau _d = 310$\,s and $Z = 55 - 100$ are consistent with
 the data in~\cite{Boukany2009n2}; with $\eta \approx 1$Pa\,s and $G
 \approx 7\times10^3$Pa~\cite{Tapadia2003} we find $\epsilon \approx
 10^{-7}$; for numerical stability we use $\epsilon = 10^{-4}$. For $L
 = 1\,$mm, $\rho \approx 10^3\,\textrm{kg m}^{-3}$ gives $\hat{\rho}
 \approx 10^{-10}$ and we use $\hat{\mathcal{D}} =
 10^{-5}$~\cite{Adams2008}. Spatial derivatives are discretized using
a semi-implicit central finite difference scheme.  For a time step
 $\delta \hat{t} = 10^{-6}$ and 1000 spatial mesh points the maximum
 velocity in the fracture and time to fracture converge within a few percent.

 We infer (in)stability by considering the evolution of perturbations to the uniform solution to Eq.\eqref{eq2} $\mathbf{s}(t) \equiv [\Delta_{xx}, \Delta_{xy}, \Delta_{yy}](t)$, where $\mathbf{\Delta} = \mathbf{W} - \mathbf{I}$, with initial conditions $\mathbf{s}(0) = [0, 0, 0]$ and imposed uniform shear rate $\hat{\dot{\gamma}}$. 
At some time $t_0$ we impose an inhomogeneous perturbation $\delta\mathbf{u}(y,t_0) = [\delta\hat{\dot{\gamma}}, \delta\Delta_{xx}, \delta\Delta_{xy}, \delta\Delta_{yy}](y,t_0) = \sum_k{\delta\mathbf{u}_k(t_0)\exp{(iky)}}$. The full dynamics is thus given by $\mathbf{u}(y,t;t_0) = [\hat{\dot{\gamma}}, \mathbf{s}](t_0) + \delta\mathbf{u}(y,t-t_0)$. The perturbation $\delta\mathbf{u} $ evolves for small times $t - t_0$ according to the dynamics given by linearizing Eqs.(\ref{eq3},\ref{eq2}): $\delta\dot{\mathbf{u}}_k(t - t_0) = \mathsf{M}_k(\mathbf{s}(t_0))\delta\mathbf{u}_k(t - t_0)$. The growth or decay of this perturbation at early times  indicates whether the perturbation can induce `fracture' after shearing is stopped at $t_0$. The perturbation will grow after $t_0$ when the largest real part $\omega_{max}$ of the spectrum of eigenvalues of $\mathsf{M}_k$ is positive.

 To capture the behaviour reported in~\cite{Boukany2009n2}, we
 consider a fluid with non-monotonic constitutive behaviour, $\beta =
 0$ [solid line in Fig.~\ref{fig1}(a)], and use $Z=72$ (consistent
 with \cite{Boukany2009n2}); this leads to shear banding and a stress
 plateau in steady state [dashes in Fig.~\ref{fig1}(a)]
 \cite{Spenley1993}. We initialize Eq.~\eqref{eq2} with random
 perturbations $\delta\mathbf{u}(0,y) =
 \xi\sum_{n=1}^5(\mathbf{A}_n/n^2)\cos n\pi y$, $A_{ni}\in [-1,1]$, where $i$ are the $4$ components of $\mathbf{A}_n$; here,  $\xi $
 sets the scale of the perturbation. The penalty $1/n^2$
 arises because high wavenumbers $n$ should be suppressed by both
 spatial gradients in $\mathbf{W}$ and by the slow dynamics of long
 wavelength velocity fluctuations that induce perturbations upon
 sample loading (for example). We use $\xi=0.01$, consistent with the
 scale of typical thermal fluctuations in $\mathbf{W}$
 \cite{adams2011}.

 Perturbations can grow if the fluid becomes
 unstable~\cite{Marrucci1983,adams2011,moorcroft2013,Supplementary}. For 34\% of 300 sets of randomly chosen $\mathbf{A}_n$, the resulting velocity profiles were similar to those
 reported in~\cite{Boukany2009n2}. Using initial conditions that
 produce the experimentally observed velocity profile, we simulate
 examples reported in~\cite{Boukany2009n2}. The green squares in Fig.~\ref{fig1}(a) are the overshoot stresses at different shear rates, and the stresses at $t_0$ for the three cases studied are indicated as I, II and III. For times $t_0$ later than the time at which the start-up stress is given by the  blue diamonds, the perturbation $\delta\mathbf{u}$ grows exponentially upon shear cessation. This is where we infer instability.

\textbf{\textit{Case I--}} For
      $\langle\dot{\gamma}\rangle\tau_R\approx 1$ and $\gamma_0 >
      \gamma _{ov}$ (the overshoot strain), we impose $\langle\hat{\dot{\gamma}}\rangle
= 200$ ($\langle{\dot{\gamma}}\rangle\tau_R =0.93$) for 
$\gamma_0=2.5$. Immediately before cessation at $t_0^-$, the velocity
profile is imperceptibly inhomogeneous [Fig.~\ref{fig1}(b)], while at
$t_0^+$ the fluid has stopped with a slight inhomogeneity induced by the
perturbation [Fig.~\ref{fig1}(d)]. Some stress then quickly relaxes due to stretch relaxation in a time $
t_s\simeq 7\tau_R$ [Fig.~\ref{fig1}(cfg)]; followed by an induction time $ t_i\simeq 30\tau_R$ with relaxation due to reptation [blue circles in Figs.~\ref{fig1}(c,f)].  The perturbation slowly grows during $t_i$ and
localizes, leading to a `fracture' plane at which the fluid shears very rapidly [Figs.~\ref{fig1}(de)] and a sizeable stretch
$\textrm{Tr}\,\mathbf{\Delta}$ is induced [Fig.~\ref{fig1}(g)]. The stress relaxes quickly during this localization in a time $ t_f\simeq 15\tau_R$ [Figs.~\ref{fig1}(cf)]. Thereafter it relaxes like a quiescent melt with a small initial
strain $\gamma_0=0.2$ [Fig.~\ref{fig1}(c)].  Since the boundaries are fixed,
positive shear strain within the slip layer is balanced by opposing
recoil in the still-entangled outer regions [\textit{e.g.}
Fig.~\ref{fig1}(e) for $t/\tau_d>0.15$].  Without an initial
perturbation only quiescent relaxation obtains [solid line of
Fig.~\ref{fig1}(c)]. The velocity profiles [Fig.~\ref{fig1}(d)] are
consistent with Fig.~1 of~\cite{Boukany2009n2} (which has an induction
time $t_i\approx 5\tau_R$).

\textbf{\textit{Stability--}} Fig.~\ref{fig1}(ac) suggests that the
material is unstable ($\omega_{\textrm{max}}>0$) from well before the
stress overshoot until shear cessation.  To understand this
instability, we turn to the Marrucci-Grizzuti (MG) observation that
for strain $\gamma_0\agt2.1$ the elastic energy function $F(\gamma)$
for the DE model has a negative effective shear modulus ${\cal
  A\/}\equiv\partial^2F/\partial\gamma^2<0$ \cite{Marrucci1983}, which
heralds instability.  MG predicted elastic instability for a step
strain, for
\begin{equation}
  {\cal A\/}^{\textrm{eff}}\equiv\mu(t_0+t_s)\left.\frac{\partial^2
      F}{\partial\gamma^2}\right|_{\gamma_0} +
  \left[1-\mu\left(t_0+t_s\right)\right]\left.\frac{\partial^2
      F}{\partial\gamma^2}\right|_{0}<0, 
\end{equation}
where $\mu(t)$ is the fraction of unrelaxed material. The
elastic limit $\dot{\gamma}\tau_d\gg1$ gives ${\cal
  A\/}^{\textrm{eff}}\simeq\partial
T_{xy}/\partial\gamma=\dot{\gamma}^{-1}\partial T_{xy}/\partial t<0$
\cite{Marrucci1983,adams2011,moorcroft2013,Adams2009b}, which
coincides with the stress overshoot.    

The anisotropy of the polymer conformation tensor $\mathbf{W}$ defines
$\mu\equiv\left|\lambda_1-\lambda_2\right|/\left|\lambda_1+\lambda_2\right|$,
where $\lambda_i$ are the eigenvalues of $\mathbf{W}$ in the plane containing the velocity gradient and flow directions \footnote{The definition of unrelaxed segments $\mu(t)$ matches
  the linear relaxation function $G(t)\equiv \lim_{\gamma _0\rightarrow 0}T_{x,y}(t,\gamma)/\gamma _0$,
  as does the
  equivalent function used by Marrucci and Grizzuti for the DE model
  \cite{Marrucci1983}}. For a homogeneous initial condition $\mu(t)$ relaxes homogeneously to zero, while an
inhomogeneous initial condition initiates instability and an
inhomogeneous $\mu(y,t)$ [Fig.~\ref{fig2}(d)].    

Fig.~\ref{fig2}(bc) shows the spatial profiles for the strain and the
effective shear modulus ${\cal A}^{\textrm{eff}}$ after stretch
relaxation \cite{energyfunction}. The fracture region is most
unstable, so that the initial perturbation [Fig.~\ref{fig2}(a)]  can localize strain. The unstable region predicted by
the elastic limit coincides with the most unstable eigenvalue
$\omega_{\textrm{max}}$ calculated from the full dynamics, which
indicates instability before the stress overshoot is reached
[\textit{e.g.} Fig.~\ref{fig1}(a)] because of the viscous contribution
to the instability \cite{moorcroft2013}. The most unstable eigenvector is dominated by the growth of $\Delta_{xx}$~\cite{Supplementary} which enhances stretch in the flow direction.
 

\textbf{\textit{Conditions for fracture--}} A detailed study shows
that perturbations in $\Delta_{xx}$ and $\Delta_{yy}$ induce fracture
\cite{Supplementary}. The step strain $\gamma_0$ advects the initial
perturbation into a shear component of the polymer strain
[\textit{e.g.} $W_{xy}(y,t_0)\simeq\gamma_0(1+\Delta_{yy}(y,0))$],
which generates an inhomogeneous shear rate
$\delta\hat{\dot{\gamma}}(y,t_0^+)\simeq-\gamma_0\Delta_{yy}(y,0)/\epsilon$
immediately after cessation of flow to maintain $\nabla\cdot
\mathbf{T}\simeq0$. Although general perturbations are complex
[Fig.~\ref{fig2}(a)] \cite{Supplementary}, a local maximum in the \textit{polymeric strain} $\gamma$
defines the position with the most negative effective shear modulus
${\cal A}_{\textrm{eff}}<0$ and the fastest growth rate $\omega_{max}$ [Fig.~\ref{fig2}(c)]
\cite{energyfunction}, and thus the fracture position.

The subsequent evolution resembles spinodal decomposition of a
conserved quantity, since the total strain $\gamma_0$ is fixed. The
strain in the most unstable region grows while that in the less
unstable regions decreases. This leads to recoil and a sharpening of the deformation
around the most unstable position, which can then fracture if the initial amplitude grows quickly enough compared to the 
overall relaxation due to reptation. Significant convected
constraint release suppresses fracture because of the enhanced
relaxation.

\textbf{\textit{Character of Fracture--}} A larger strain leads to a
less dramatic fracture [Fig.~\ref{fig3}(ad)] because
the total stress has passed the overshoot and decreased, hence
releasing less stress into the fracture; however the larger molecular strain
$W_{xy}$ leads to a faster growing instability, which 
is consistent with Fig.~8 of~\cite{Boukany2009n2}. Alternatively, for
a higher imposed strain rate and $t_0$ beyond the overshoot the stretch-dominated response leaves less orientational stress and molecular
strain after stretch relaxation, so that fracture takes longer to
develop \cite{moorcroft2013}  

In \textbf{\textit{Case
      II}} ($\langle\hat{\dot{\gamma}}\rangle = 900$,
  $\langle\dot{\gamma}\rangle\tau_R$ $= 4.2$), the shear rate is large
  but the strain $\gamma_0=2.5$ is slightly less than the overshoot
  strain $\gamma_{ov}$ [Fig.~\ref{fig3}(bc)].  The velocity profiles
  are consistent with Fig.~2
  of~\cite{Boukany2009n2}. 
  Because the growth rate
  $\omega_{\textrm{max}}$ is so rapid for the high shear rate, the
  smaller strain can effect the necessary large growth of the
  instability. In this case the induction time and velocity profiles are similar
    to Case I. In {\textbf{\textit{Case III}}
    ($\langle\hat{\dot{\gamma}}\rangle = 10, \langle\dot{\gamma}\rangle\tau_R = 0.046$) the shear rate is
    relatively small [Fig.~\ref{fig3}(ef)], and `fracture' and recoil are very
    weak due to the small growth  
    rate. The stress response due to the inhomogeneity is almost negligible compared to that of an unperturbed initial condition. The weak recoil agrees with
    Fig.~7 of~\cite{Boukany2009n2}.}

    Fig.~6 of~\cite{Boukany2009n2} demonstrated that, for sub-overshoot strains, higher shear rates lead to longer induction times; while our calculations predict shorter induction times  because of the faster growing instability \cite{Supplementary}. We cannot explain this discrepancy. 

    \textbf{\textit{Conclusion--}} We have shown that the
    ``fracture'' seen in recent step strain experiments on polymeric
    liquids \cite{Boukany2009n2,FangJOR2011} could result from an
    underlying elastic instability in the DE model, whose signature is stress overshoot
    during rapid
    startup~\cite{Marrucci1983,*Morrison1992,Adams2009,*CaoLikhtmanPRL2012,Moorcroft2011}. Once
    stretch degrees of freedom have relaxed, the deformed melt is
    elastically unstable so that small inhomogeneities grow into
    plastic strain (shear flow) in the most unstable regions. If this
    instability grows fast enough compared to reptation then a
    dramatic fracture can result.  The perturbation's shape and amplitude control
    whether fracture occurs.

    In related works, Manning \textit{et al.} studied a shear-transformation-zone model of an amorphous
    solid \cite{ManLan2007PSNSMP,*Manning2009rate}, demonstrating  plastic  yield
within a  fluid shear band (or fracture) during startup of
    shear flow; while a shear-dilation coupling has been shown to lead to fracture in glass-forming materials \cite{furukawa2009inhomogeneous}. In the rubbery polymer liquid considered here the instability is purely constitutive: shearing leads to a decreased stress as chains are oriented along the flow direction, and the resulting fluid is mechanically unstable. 

    Boukany \textit{et al.} suggested that the fracture demands new
    physics \cite{Boukany2009n2}. Certainly current tube models are
    incomplete \cite{likhtman2009whither}.  However, our calculations
    are reasonable if spatial features are smooth on length scales
    greater than the tube diameter $a\simeq3-4\,\textrm{nm}$. For a
    gap of $1\,\textrm{mm}$, the fracture width $\delta x\simeq0.05$
    corresponds to a thickness of order $50\,\mu\textrm{m}$, which is
    consistent with the dimension $\leq 40\,\mu\textrm{m}$ reported in
    Ref.~\cite{Boukany2009n2}. Thus, higher experimental resolution will
    determine whether or not the continuum nature of the tube
    model is adequate.

    \textbf{\textit{Acknowledgments--}} This study was funded by the
    EU ITN DYNACOP. We thank Robyn Moorcroft, Suzanne Fielding, and
    Scott Milner for helpful advice.
    
    
\widetext
\begin{center}
\textbf{Apparent Fracture in Polymeric Fluids under Step Shear: Supplementary Information}
\end{center}

\section{Calculations}
\subsection{Step-Strain Calculations for Different Initial Conditions}

The starting point for the calculations is the diffusive Rolie-Poly (DRP) model, given by~\cite{Likhtman2003,*Adams2009}
\begin{align}
\frac{d\mathbf{W}}{dt} & = \mbox{\boldmath{$\kappa$}} \cdot \mathbf{W} + \mathbf{W}\cdot\mbox{\boldmath{$\kappa$}}^T - \frac{1}{\tau
  _d}(\mathbf{W} - \mathbf{I}) - \frac{2\left(1 -\sqrt{\frac{3}{\textrm{Tr}\mathbf{W}}}\right)}{\tau_R}\left(\mathbf{W} +
  \beta\left(\frac{\textrm{Tr}\mathbf{W}}{3}\right)^{\delta}(\mathbf{W}
  - \mathbf{I})\right) + \mathcal{D}\nabla ^2 \mathbf{W},
  \label{eq1s}
\end{align}
where $\mathbf{W}$ is a polymer strain, $\kappa _{\alpha\beta} = \partial _{\alpha}v_{\beta}$, $\mathbf{v}$ is the fluid velocity, $\tau _d$ and $\tau _R$ are the reptation and stretch relaxation times respectively, $\mathbf{I}$ is the identity tensor, $\beta$ measures the amount of convective constraint release in the system, $\delta$ is a fitting parameter and $\mathcal{D}$ is stress diffusion constant.  

We use the Cartesian coordinate system (for the case of simple shear flow where the fluid is placed between two infinite parallel plates of separation $L$) where $\mathbf{\hat{y}}$ is the velocity gradient direction and $\mathbf{\hat{x}}$ is the flow direction, $\mathbf{v} = v_x(t,y)\mathbf{\hat{x}}$ and $\mathbf{W} = \mathbf{W}(t,y)$. Substitution into Eq.~\ref{eq1s} with $\mathbf{W} = \mathbf{\Delta} + \mathbf{I}$ gives
\begin{subequations}
\begin{align}
\frac{\partial \Delta _{xx}}{\partial t} & = 2\Delta _{xy}\widehat{\dot{\gamma}} - \Delta_{xx} - \frac{2\tau _d}{\tau _R}[1 - A]\left[\left(\beta A + 1\right)\Delta _{xx} + 1\right]   + \widehat{\mathcal{D}}\frac{\partial ^2\Delta _{xx}}{\partial y^2} \\
\frac{\partial \Delta _{xy}}{\partial t} & = \widehat{\dot{\gamma}} + \widehat{\dot{\gamma}}\Delta_{yy} - \Delta_{xy} - \frac{2\tau _d}{\tau _R}[1 - A](\beta A + 1)\Delta _{xy} + \widehat{\mathcal{D}}\frac{\partial ^2\Delta _{xy}}{\partial y^2} \\
\frac{\partial \Delta _{yy}}{\partial t} & = - \Delta _{yy} - \frac{2\tau _d}{\tau _R}[1 - A][(\beta A + 1)\Delta _{yy} + 1] + \widehat{\mathcal{D}}\frac{\partial ^2\Delta _{yy}}{\partial y^2}  \\
A & = \left(1 + \frac{\textrm{Tr}\mathbf{\Delta}}{3}\right)^{-1/2},
\end{align}
\label{eq2s}
\end{subequations}
where $\widehat{\dot{\gamma}} = \dot{\gamma}\tau _d$, $\widehat{\mathcal{D}} = \mathcal{D}\tau _d/L^2$~\footnote{The time $t$ and spatial variable $y$ have been made dimensionless as $\hat{t} = t/\tau_d$ and $\hat{y} = y/L$ respectively. However, for simplicity the quantities $\hat{t}$ and $\hat{y}$ have been written as $t$ and $y$ in Eq.~\ref{eq2s}}. The total stress $\mathbf{T}$ is then obtained from $\mathbf{W}$ and a Newtonian solvent of viscosity $\eta$ as 
\begin{equation}
\mathbf{T} = G\mathbf{W} + \eta (\mbox{\boldmath{$\kappa$}} +
\mbox{\boldmath{$\kappa$}}^T) - p\mathbf{I}, 
\label{eq3s}
\end{equation}
where $G$ is the plateau modulus and $p$ is pressure, this gives the total shear stress as
\begin{equation}
T_{xy} = G\Delta _{xy} + \eta\dot{\gamma}.
\label{eq4s}
\end{equation}
\begin{figure}[h]
\begin{center}
{\includegraphics[width=0.9\textwidth]{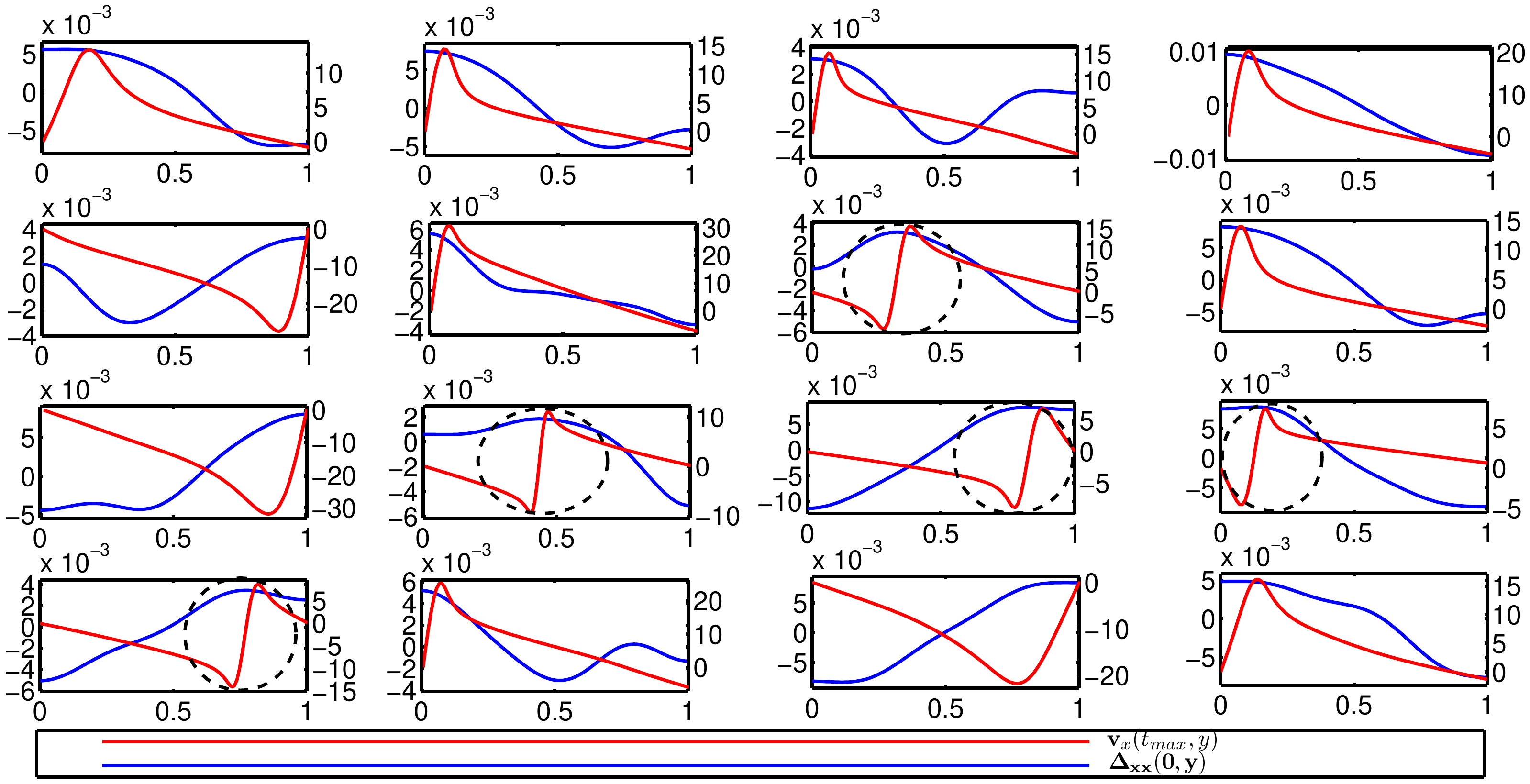}}
\end{center}
\caption{Recoil or `fracture' for different random initial conditions upon perturbing $\Delta_{xx}$.
        In all cases, the blue line is the perturbation and the red line is the velocity profile when both $\mu_{\pm}$ reach their extrema together (for `fracture') or $\mu_{\pm}$ reach their extrema separately (for recoil without `fracture'). The `fracture' profiles are indicated by the dashed circles.
Left  axis: perturbation; 
Right  axis: velocity.} 
 \label{fig1s}
 \end{figure}
\begin{figure}[h]
\begin{center}
{\includegraphics[width=0.9\textwidth]{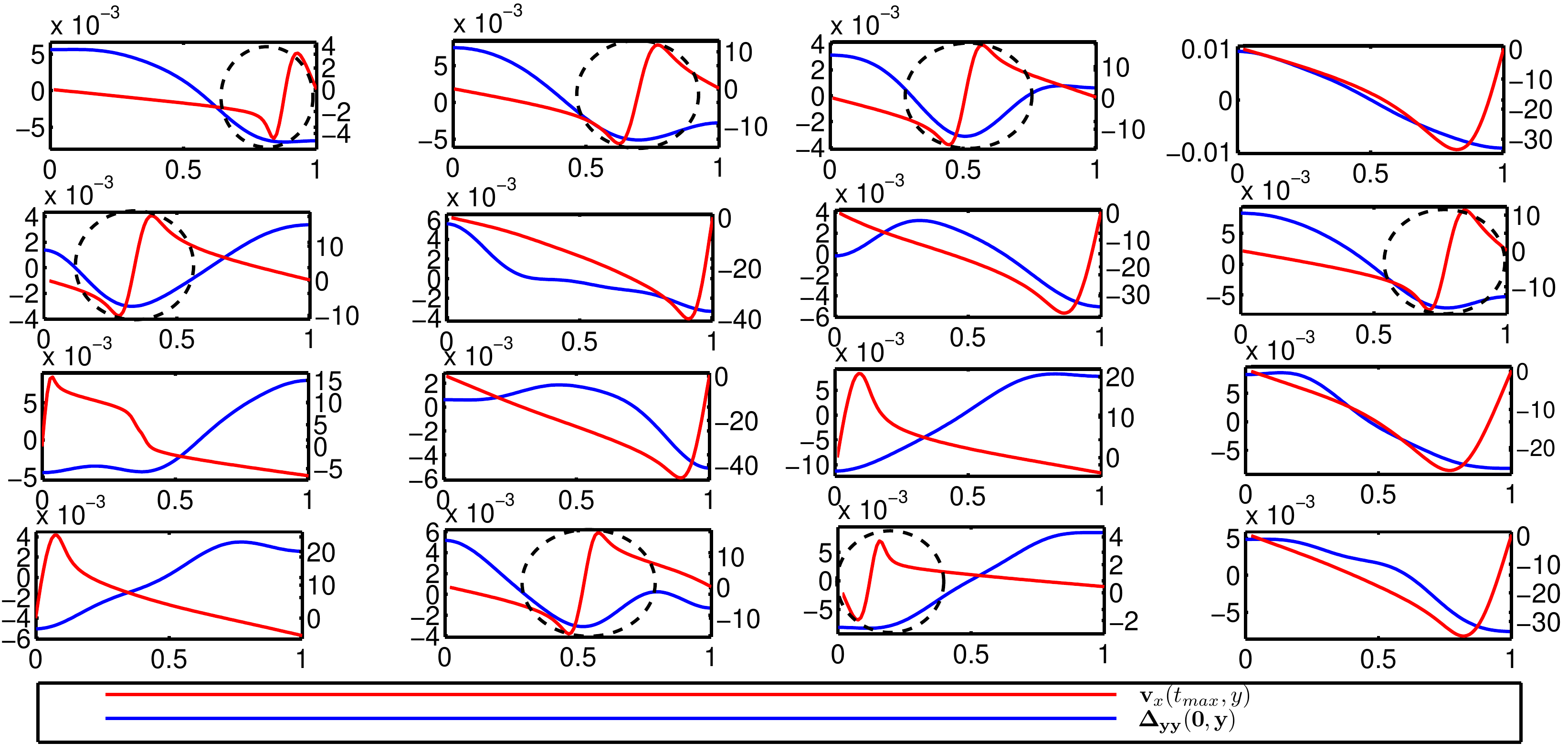}}
\end{center}
\caption{Same as Figure~\ref{fig1s} but with perturbation to $\Delta_{yy}$.} 
 \label{fig2s}
 \end{figure}
 \begin{figure}[htb]
\begin{center}
{\includegraphics[width=0.9\textwidth]{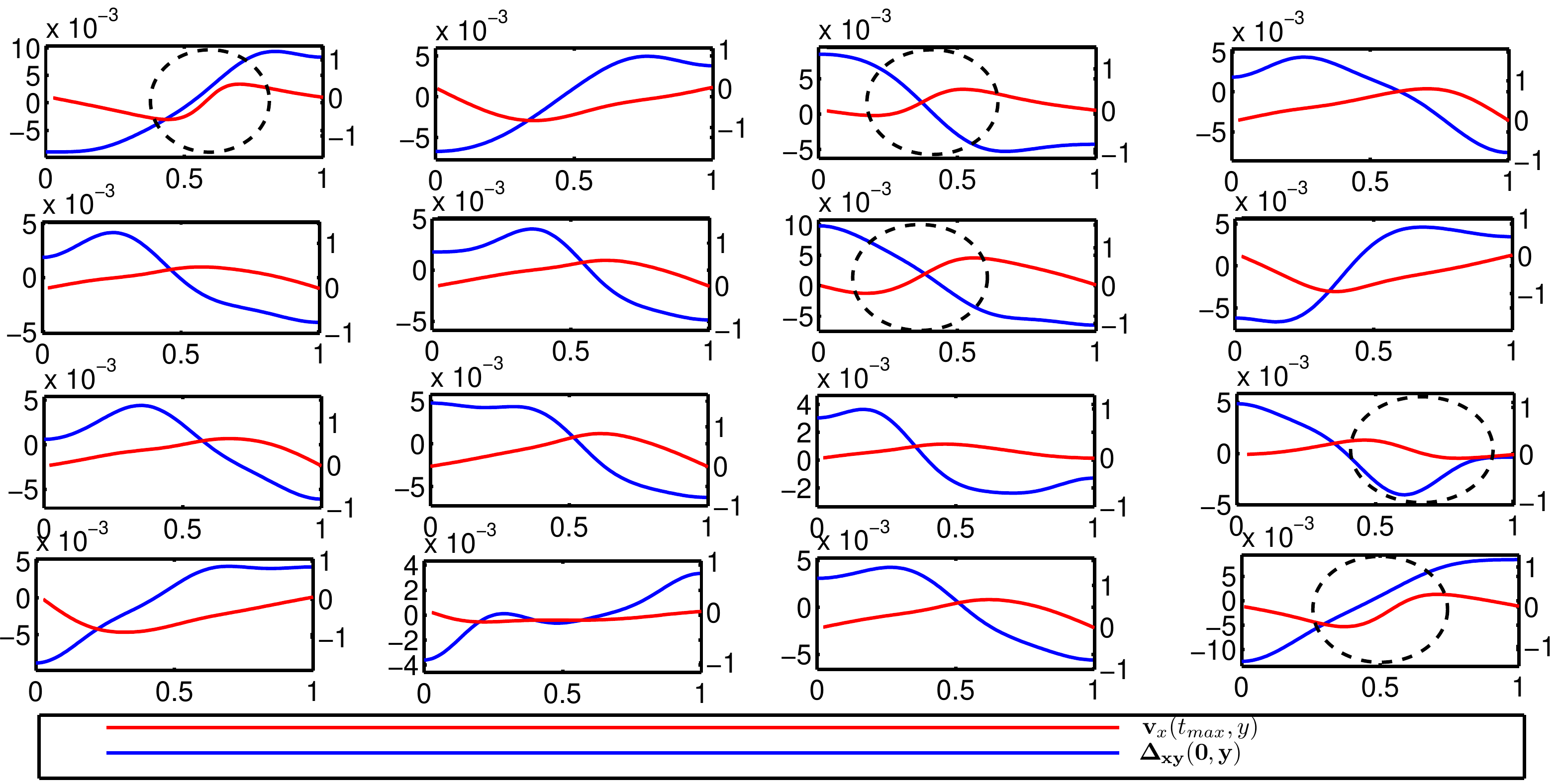}}
\end{center}
\caption{Perturbation of $\Delta_{xy}$. Only a weak recoil or a weak sign of `fracture' is seen in this case. The blue and  red lines have the same meaning as in Fig.~\ref{fig1s}. The `weak fracture' profiles are indicated by the dashed circles. Left axis: perturbation; 
Right axis: velocity.} 
 \label{fig3s}
 \end{figure}
 \begin{figure}[htb]
\begin{center}
{\includegraphics[width=0.9\textwidth]{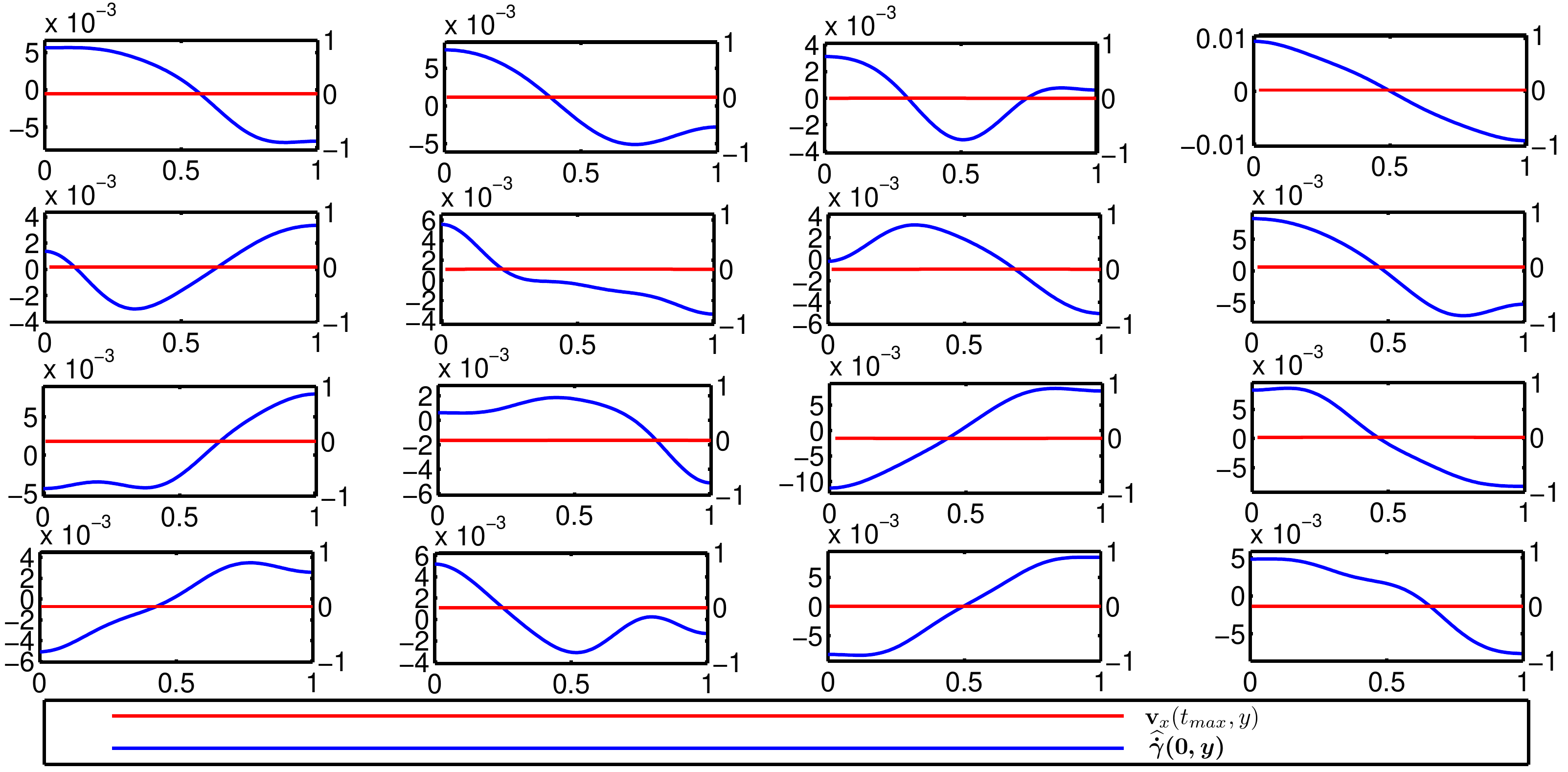}}
\end{center}
\caption{Same as Figure~\ref{fig3s} but with perturbation to $\hat{\dot{\gamma}}$.} 
 \label{fig4s}
 \end{figure}
 \begin{figure}[htb]
\begin{center}
{\includegraphics[width=0.9\textwidth]{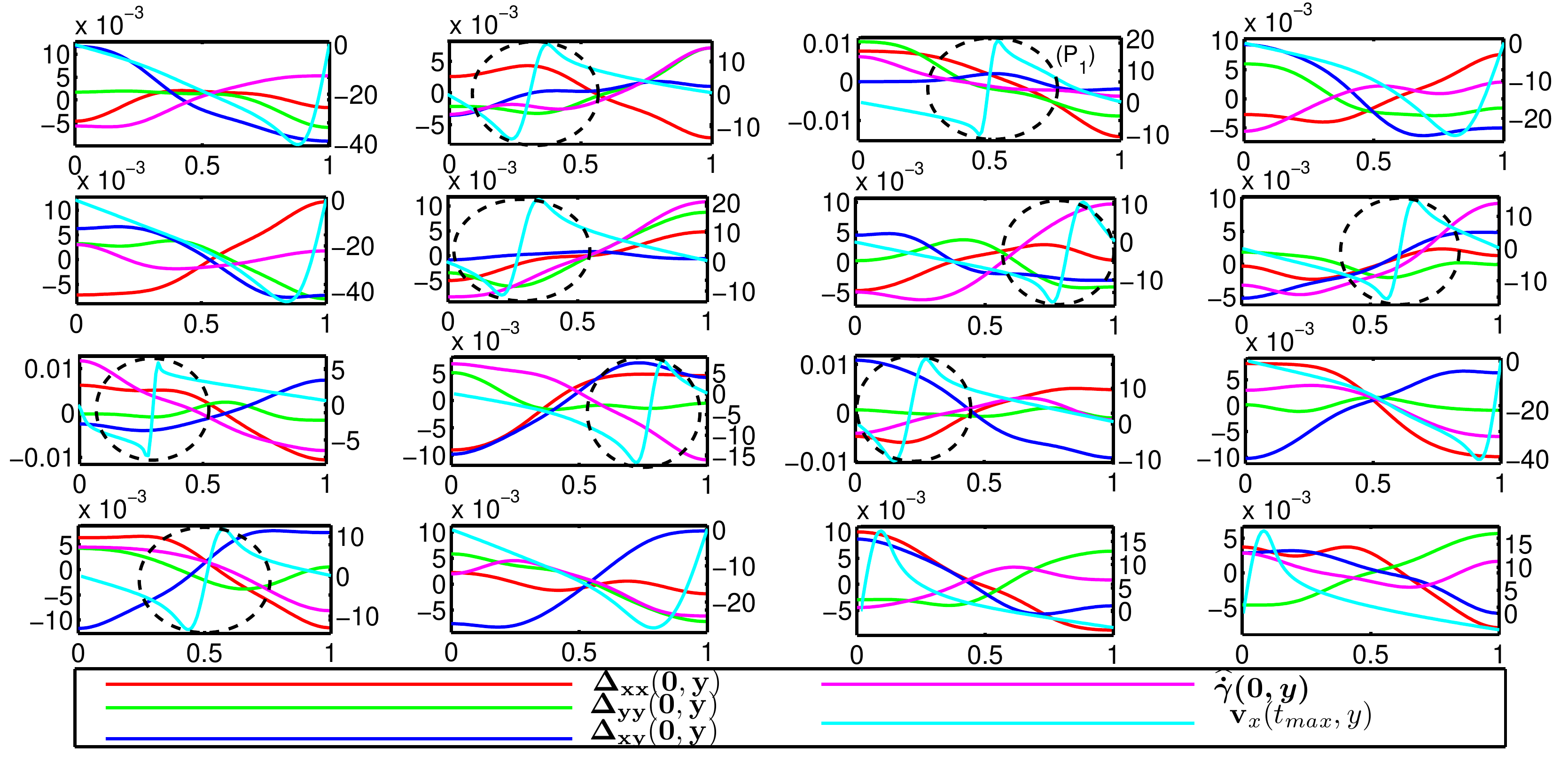}}
\end{center}
\caption{Recoil or `fracture' upon perturbing all components, with each component receiving a separate random perturbation.
Red line: perturbation to $\Delta_{xx}$.
Green line: perturbation to $\Delta_{yy}$.
Blue line: perturbation to $\Delta_{xy}$. 
Magenta line: perturbation to $\dot{\gamma}$.
Cyan line: recoil or `fracture' velocity profile $v$.
Left axis: perturbation; 
Right axis: velocity. In all cases, the `fracture' profiles are indicated with the dashed circles.}
 \label{fig5s}
 \end{figure}
 \begin{figure}[htb]
\begin{center}
{\includegraphics[width=8.3truecm]{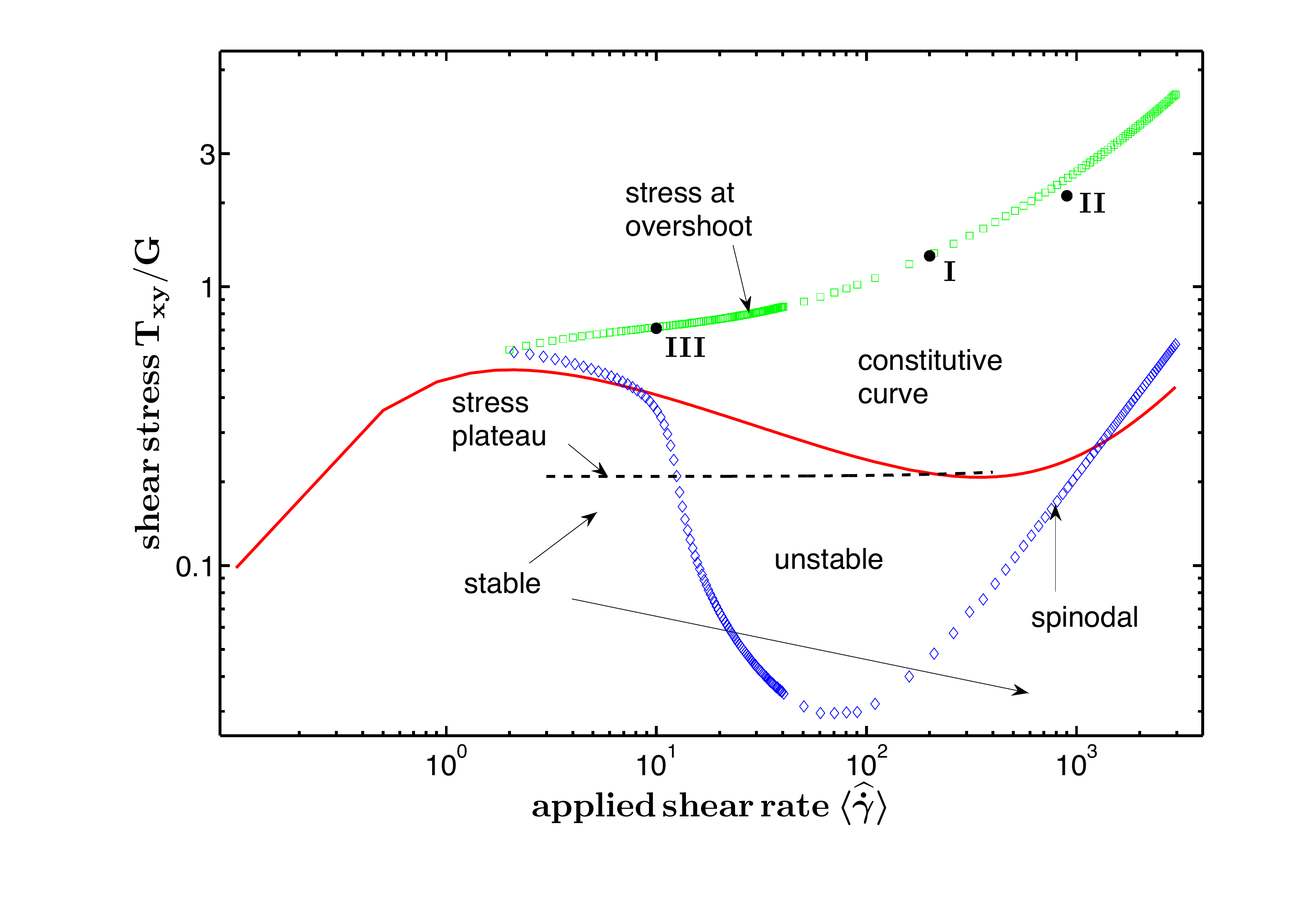}}
\end{center}
\caption{Spinodal (shear stresses at which the perturbation imposed on the base state at $t_0$ grow), constitutive curve, stress plateau and the overshoot stress for the DRP model. The stresses at shear cessation for the three Cases discussed in the manuscript are indicated as I, II and III. Parameters: $\beta = 0$, $Z = \tau _d/(3\tau _R) = 72$.}
 \label{fig6s}
 \end{figure}

To capture the behaviour reported in~\cite{Boukany2009n2}, we initialize Eq.~\ref{eq2s} with random perturbations of the form
\begin{equation}
\delta\mathbf{u}(0,y) = \xi\sum _{n=1}^5(\mathbf{A}_n/n^2)\cos (n\pi y),
\label{eq5s}
\end{equation}
where $\mathbf{u} \equiv [\widehat{\dot{\gamma}}, \Delta _{xx}, \Delta _{xy}, \Delta _{yy}]$. The amplitudes $A_{ni}$ which are the components of vector $\mathbf{A}$ are chosen randomly within $[-1,1]$. The index $i = 1,2,3,4$ corresponds to each of the quantities $[\widehat{\dot{\gamma}}, \Delta _{xx}, \Delta _{xy}, \Delta _{yy}]$
The parameter $\xi = 0.01$ sets the overall scale of amplitude and a cosine series was chosen since it satisfies the boundary condition imposed on $\mathbf{\Delta}$. Using more modes does not change the resultant perturbation significantly due to the $1/n^2$ penalty on the amplitudes. Each component of $\mathbf{u}$ is initially perturbed separately using different random perturbations. Then all components of $\mathbf{u}$ are perturbed together with each quantity receiving a separate random perturbation. Sample results from these simulations are shown in Figs.~\ref{fig1s} to~\ref{fig5s}.

In all calculations reported here, the parameters were set as $Z = \tau _d/(3\tau _R) = 72$, $\widehat{\mathcal{D}} = 10^{-5}$, $\epsilon = \eta/(G\tau _d) = 10^{-4}$, $\beta = 0$ and for stability analysis, $\hat{\rho} = 10^{-10}$. To determine if fracture has occurred or not, consider the `velocity moments' $\mu _{v\pm}$, defined by
\begin{equation}
\mu _{v\pm} = \sum _i{v_iH(\pm v_i)},
\label{eq6s}
\end{equation}
where the sum is over all spatial positions $y_i$ and $H$ is the Heaviside step function. If both positive moment $\mu _{v+}$ and negative moment $\mu _{v\_}$ occur together at any time during stress relaxation after shear cessation, then we say that `fracture' has occurred, otherwise there is no fracture. The velocity profiles shown in Figs.~\ref{fig1s} to~\ref{fig5s} occur at the time when both $\mu _{v+}$ and $\mu _{v\_}$ reach their extrema for the case of fracture. When there is no fracture, the velocity profiles are shown when either $\mu _{v+}$ reaches its maximum or $\mu _{v\_}$ reaches its minimum. When fracture occurs, the position of the fracture plane depends on the shape of the specific perturbation. The stress relaxation is independent of the position of the fracture plane, as in the experiments of~\cite{Boukany2009n2} (section III A).  
 
 In about 34\% of 300 simulations where $\Delta _{xx}$, $\Delta _{yy}$, $\Delta _{xy}$ and $\widehat{\dot{\gamma}}$ are all perturbed simultaneously, the resultant velocity profiles resemble the type reported in~\cite{Boukany2009n2}. The calculations in the manuscript use a set of initial conditions that give a fracture with all quantities perturbed, such as subfigure $\textrm{P}_1$ in Fig.~\ref{fig5s}. 

\subsection{\textit{Linear stability analysis}} 
Linear stability analysis is carried out by considering the stability of a homogeneous base state $\mathbf{s}(t)$ to fluctuations. During the evolution of the base state $\mathbf{s}(t)$, a perturbation $\delta\mathbf{u}(t,y) = [\delta\Delta _{xx},\delta\Delta _{xy},\delta\Delta _{yy},\delta\widehat{\dot{\gamma}}](t,y)$ is introduced at some time $t_0$. Subsequent evolution of the perturbation is then given by
\begin{equation}
\mathbf{u}(t;t_0,y) = [\hat{\dot{\gamma}}, \mathbf{s}](t_0) + \delta\mathbf{u}(t-t_0,y).
\label{eq7s}
\end{equation}
If the perturbation grows at early times after shear cessation at $t_0$, then it may be able to induce `fracture' at later times. The homogeneous base state $\mathbf{s}(t) = [\widehat{\overline{\dot{\gamma}}}, \overline{\Delta}_{xx}, \overline{\Delta}_{xy}, \overline{\Delta}_{yy}]$ is obtained by solving
\begin{subequations}
\begin{align}
\partial _t\overline{\Delta} _{xx} = & 2\overline{\Delta} _{xy}\widehat{\overline{\dot{\gamma}}} - \overline{\Delta}_{xx} - \frac{2\tau _d}{\tau _R}\left[1 - \overline{A}\right]\left[\left(\beta \overline{A} + 1\right)\overline{\Delta} _{xx} + 1\right] \\
\partial _t\overline{\Delta} _{xy} = & \widehat{\overline{\dot{\gamma}}} + \widehat{\overline{\dot{\gamma}}}\,\overline{\Delta}_{yy} - \overline{\Delta}_{xy} - \frac{2\tau _d}{\tau _R}\left[1 - \overline{A}\right]\left(\beta \overline{A} + 1\right)\overline{\Delta} _{xy}\\
\partial _t\overline{\Delta} _{yy}= & - \overline{\Delta} _{yy} - \frac{2\tau _d}{\tau _R}\left[1 - \overline{A}\right]\left[\left(\beta \overline{A} + 1\right)\overline{\Delta} _{yy} + 1\right]  \\
\overline{A} = & \left(1 + \frac{\textrm{Tr}\mathbf{\overline{\Delta}}}{3}\right)^{-1/2}.
\end{align}
\label{eq8s}
\end{subequations}
The perturbation $\delta\mathbf{u}(t,y)$ consists of fluctuations in the velocity gradient direction of the form
\begin{equation}
\delta\mathbf{u}(t,y) = \sum _k\delta\mathbf{u}_k(t)\exp (iky)\quad t\geq t_0.
\label{eq9s}
\end{equation}
Substituting Eq.~\ref{eq7s} into Eq.~\ref{eq2s} and the momentum equation
\begin{equation}
  \rho\frac{d\mathbf{v}}{dt} \equiv \rho\left[\frac{\partial}{\partial t}  +
    \left(\mathbf{v}\cdot \nabla\right)\right]\mathbf{v} = \nabla
  \cdot\mathbf{T}, 
\label{eq10s}
\end{equation}
where $\rho$ is the fluid density, gives
\begin{subequations}
\begin{align}
\partial _t\delta\Delta _{xx,k}(t) = & \left[\frac{\tau _d}{3\tau _R}\overline{\Delta}_{xx}\left(\beta - 1\right)\overline{A}^3 -1 - \frac{\tau _d}{3\tau _R}\overline{A}^3 - 2\frac{\tau _d}{\tau _R}\left[1 + \left(\beta -  1\right)\overline{A}\right] - \frac{2}{3}\beta\overline{\Delta}_{xx}\overline{A}^4 + 2\beta\overline{A}^2 \right.\nonumber\\
& \left. - k^2\widehat{\mathcal{D}}\right]\delta\Delta _{xx,k}(t) \nonumber\\
& + 2\widehat{\overline{\dot{\gamma}}}\delta\Delta _{xy,k}(t)\nonumber\\
& + \left[\left(\beta - 1\right)\frac{\tau _d}{3\tau _R}\overline{\Delta}_{xx}\overline{A}^3 - \frac{\tau _d}{3\tau _R}\overline{A}^3 - \frac{2}{3}\beta\overline{\Delta}_{xx}\overline{A}^4\right]\delta\Delta _{yy,k}(t)\nonumber\\
 & + 2\overline{\Delta}_{xy}\delta\widehat{\dot{\gamma}}_{k}(t) \\
\partial _t\delta\Delta _{xy,k}(t) = & \left[\left(\beta - 1\right)\frac{\tau _d}{3\tau _R}\overline{\Delta}_{xy}\overline{A}^3 - \frac{2}{3}\beta\frac{\tau _d}{\tau _R}\overline{A}^4\overline{\Delta}_{xy}\right]\delta\Delta _{xx,k}(t)\nonumber\\
 & + \left[2\beta\frac{\tau _d}{\tau _R}\overline{A}^2 - 1 - 2\frac{\tau _d}{\tau _R}\left[1 + \left(\beta - 1\right)\overline{A}\right] - k^2\widehat{\mathcal{D}}\right]\delta\Delta _{xy,k}(t) \nonumber\\
 & + \left[\widehat{\overline{\dot{\gamma}}} + \left(\beta - 1\right)\frac{\tau _d}{3\tau _R}\overline{\Delta}_{xy}\overline{A}^3 - \frac{2}{3}\beta\frac{\tau _d}{\tau _R}\overline{\Delta}_{xy}\overline{A}^4\right]\delta\Delta _{yy,k}(t) \nonumber\\
 & + \left[1 + \overline{\Delta}_{yy}\right]\delta\widehat{\dot{\gamma}}_{k}(t) \\
 \partial _t\delta\Delta _{yy,k}(t) = & \left[\left(\beta - 1\right)\frac{\tau _d}{3\tau _R}\overline{\Delta}_{yy}\overline{A}^3 - \frac{\tau _d}{3\tau _R}\overline{A}^3 - \frac{2}{3}\beta\frac{\tau _d}{\tau _R}\overline{\Delta}_{yy}\overline{A}^4\right]\delta\Delta _{xx,k}(t) \nonumber\\
 & + \left[\frac{\tau _d}{3\tau _R}\left(\beta - 1\right)\overline{\Delta}_{yy}\overline{A}^3 - 1 - \frac{\tau _d}{3\tau _R}\overline{A}^3 - 2\frac{\tau _d}{\tau _R}\left[1 + \left(\beta - 1\right)\overline{A}\right] - \frac{2}{3}\beta\frac{\tau _d}{\tau _R}\overline{\Delta}_{yy}\overline{A}^4 \right.\nonumber\\
 & \left. + 2\beta\frac{\tau _d}{\tau _R}\overline{A}^2 - k^2\widehat{\mathcal{D}}\right]\delta\Delta _{yy,k}(t)\\
\partial _t\delta\widehat{\dot{\gamma}}_{k}(t) = & - \frac{k^2}{\widehat{\rho}}\delta\Delta _{xy,k}(t) - \frac{k^2\epsilon}{\widehat{\rho}}\delta\widehat{\dot{\gamma}}_{k}(t) \\
\epsilon = & \frac{\eta}{G\tau _d} \\
\widehat{\rho} = & \frac{\rho L^2}{G\tau _d^2},
\end{align}
\label{eq11s}
\end{subequations}
where all nonlinear terms in $[\delta\Delta _{xx,k},\delta\Delta _{xy,k},\delta\Delta _{yy,k},\delta\widehat{\dot{\gamma}}_{k}]$ have been neglected. In the zero Reynolds number limit $\hat{\rho}\rightarrow 0$ this reduces to
\begin{subequations}
\begin{align}
\partial _t\delta\Delta _{xx,k}(t) = & \left[\frac{\tau _d}{3\tau _R}\overline{\Delta}_{xx}\left(\beta - 1\right)\overline{A}^3 -1 - \frac{\tau _d}{3\tau _R}\overline{A}^3 - 2\frac{\tau _d}{\tau _R}\left[1 + \left(\beta -  1\right)\overline{A}\right] - \frac{2}{3}\beta\overline{\Delta}_{xx}\overline{A}^4 + 2\beta\overline{A}^2 \right.\nonumber\\
& \left. - k^2\widehat{\mathcal{D}}\right]\delta\Delta _{xx,k}(t) \nonumber\\
& + 2\left[\widehat{\overline{\dot{\gamma}}} - \frac{\overline{\Delta} _{xy}}{\epsilon}\right]\delta\Delta _{xy,k}(t)\nonumber\\
& + \left[\left(\beta - 1\right)\frac{\tau _d}{3\tau _R}\overline{\Delta}_{xx}\overline{A}^3 - \frac{\tau _d}{3\tau _R}\overline{A}^3 - \frac{2}{3}\beta\overline{\Delta}_{xx}\overline{A}^4\right]\delta\Delta _{yy,k}(t) \\
\partial _t\delta\Delta _{xy,k}(t) = & \left[\left(\beta - 1\right)\frac{\tau _d}{3\tau _R}\overline{\Delta}_{xy}\overline{A}^3 - \frac{2}{3}\beta\frac{\tau _d}{\tau _R}\overline{A}^4\overline{\Delta}_{xy}\right]\delta\Delta _{xx,k}(t)\nonumber\\
 & + \left[2\beta\frac{\tau _d}{\tau _R}\overline{A}^2 - 1 - 2\frac{\tau _d}{\tau _R}\left[1 + \left(\beta - 1\right)\overline{A}\right] - \frac{1 + \overline{\Delta} _{yy}}{\epsilon} - k^2\widehat{\mathcal{D}}\right]\delta\Delta _{xy,k}(t) \nonumber\\
 & + \left[\widehat{\overline{\dot{\gamma}}} + \left(\beta - 1\right)\frac{\tau _d}{3\tau _R}\overline{\Delta}_{xy}\overline{A}^3 - \frac{2}{3}\beta\frac{\tau _d}{\tau _R}\overline{\Delta}_{xy}\overline{A}^4\right]\delta\Delta _{yy,k}(t) \\
 \partial _t\delta\Delta _{yy,k}(t) = & \left[\left(\beta - 1\right)\frac{\tau _d}{3\tau _R}\overline{\Delta}_{yy}\overline{A}^3 - \frac{\tau _d}{3\tau _R}\overline{A}^3 - \frac{2}{3}\beta\frac{\tau _d}{\tau _R}\overline{\Delta}_{yy}\overline{A}^4\right]\delta\Delta _{xx,k}(t) \nonumber\\
 & + \left[\frac{\tau _d}{3\tau _R}\left(\beta - 1\right)\overline{\Delta}_{yy}\overline{A}^3 - 1 - \frac{\tau _d}{3\tau _R}\overline{A}^3 - 2\frac{\tau _d}{\tau _R}\left[1 + \left(\beta - 1\right)\overline{A}\right] - \frac{2}{3}\beta\frac{\tau _d}{\tau _R}\overline{\Delta}_{yy}\overline{A}^4 \right.\nonumber\\
 & \left. + 2\beta\frac{\tau _d}{\tau _R}\overline{A}^2 - k^2\widehat{\mathcal{D}}\right]\delta\Delta _{yy,k}(t)
\end{align}
\label{eq12s}
\end{subequations}
This is a matrix equation of the form
\begin{equation}
\partial _t\delta\mathbf{\tilde{u}}(t) = \mathsf{M}(t_0)\cdot\delta\mathbf{\tilde{u}}(t) \quad t \geq t_0,
\label{eq13s}
\end{equation}
where $\mathbf{\tilde{u}} = [\Delta _{xx}, \Delta _{xy}, \Delta _{yy}]$.
Similar to the case described in~\cite{fielding03a}, the eigenvalues of the stability matrix $\mathsf{M}(t_0)$ determine the (in)stability of the system. We infer instability when the the largest real part of an eigenvalue just becomes positive~\cite{fielding03a}. In this situation the perturbations grow exponentially. Hence the spinodal (the shear stress at which the fluid goes unstable during startup) for the system can be constructed as shown in Fig.~\ref{fig6s}. This region of instability matches the constitutive curve, similar to the situation reported in~\cite{fielding03a}.

When the perturbation given in Eq.~\ref{eq5s} is used to initialize the system, it induces some inhomogeneity in the system. Each point in space can then be considered as a base state and the stability of each of these base states to small amplitude fluctuations is also described by the stability matrix $\mathsf{M}(t_0)$. Hence the most unstable of these base states (which is the state whose eigenvalue has the largest real part) can be determined. This approach gives insight into the behaviour of the system when the quantities $\dot{\gamma}$,  $\Delta _{xx}$, $\Delta _{xy}$ and $\Delta _{yy}$ are perturbed separately. For 15 different initial conditions that give a `fracture' profile, the eigenvector $\mathbf{\widetilde{v}}_m$ corresponding to the maximum real eigenvalue in space at the time of stretch relaxation is heavily dominated by the components $\Delta _{xx}$ and $\Delta _{yy}$. The components of $\mathbf{\widetilde{v}}_m$ for these different initial conditions are shown in Table~\ref{tab1}, where $\widetilde{v}_m^{xx}$ is the component in the flow direction, $\widetilde{v}_m^{xy}$ is the component in the shear direction and $\widetilde{v}_m^{yy}$ is the component in the velocity gradient direction.
\begin{table}
\caption{Components of most unstable eigenvector $\mathbf{\widetilde{v}}_m$ for 15 different initial conditions, for $\langle\widehat{\dot{\gamma}}\rangle = 200$, $\gamma _0 = 2.5$.}
\label{tab1}
\begin{center}
\begin{tabular}{c|c|c|c}
\hline
\hline
Initial Condition	&	$\widetilde{v}_m^{xx}$ &	$\widetilde{v}_m^{xy}$	&	$\widetilde{v}_m^{yy}$ 	\\
\hline
1	  &	0.9657	&	-0.0115	&	-0.2594		 \\
2	  &	0.9767	&	-0.0131	&	-0.2144			\\
3	  & 0.9678	&	-0.0118	&	-0.2516			\\
4	  &	0.9730	&	-0.0125	&	-0.2304			\\
5	  &	0.9632	&	-0.0112	&	-0.2687			\\
6	  &	0.9667	&	-0.0117	&	-0.2555			\\
7	  &	0.9678	&	-0.0118	&	-0.2513			\\
8	  &	0.9636	&	-0.0113	&	-0.2672			\\
9	  &	0.9739	&	-0.0127	&	-0.2266		  \\
10	&	0.9683	&	-0.0119	&	-0.2496			\\
11	&	0.9675	&	-0.0118	&	-0.2524			\\
12	&	0.9742	&	-0.0127	&	-0.2255			\\
13	&	0.9666	&	-0.0116	&	-0.2562			\\
14	&	0.9613	&	-0.0110	&	-0.2753			\\
15	&	0.9596	&	-0.0109	&	-0.2812			\\
\hline
\hline
\end{tabular}
\end{center}
\end{table}
Hence perturbing the components $\dot{\gamma}$ and $\Delta _{xy}$ separately do not induce `fracture' (as in Figs.~\ref{fig3s} and~\ref{fig4s}) as compared with perturbing the components $\Delta _{xx}$ and $\Delta _{yy}$ separately at the same amplitude (as in Figs.~\ref{fig1s} and~\ref{fig2s}).
 
\subsection{\textit{Comparison with experiment}}
The calculations in the manuscript are based on the sample SBR 250K whose rheological properties are reported in Tables 1 and 2 of~\cite{Boukany2009n2}. The rheological properties reported in Table 2 of~\cite{Boukany2009n2} were said to have been measured from linear viscoelastic measurements (see section II B of~\cite{Boukany2009n2}) but the Rouse times reported in Table 2 were estimated using $\tau _R^w = \tau_d/(M_w/M_e)$, where $\tau_d$ is the reptation time (section II B of~\cite{Boukany2009n2}). However, in our manuscript the Rouse time is calculated using $\tau _R = \tau _d/(3Z)$ (as given in section I of~\cite{Graham2003}), where $Z = M_w/M_e$ is the number of entanglements per chain. This then implies that the values of $\tau _R^w$ quoted in Table 2 of~\cite{Boukany2009n2} are larger than the values of $\tau _R$ used in our manuscript by a factor of 3.
We then present the data in~\cite{Boukany2009n2} as different cases.

\textbf{\textit{Case I: Intermediate Shear Rate, High Strain--}} Using $\langle\dot{\gamma}\rangle = 0.7\,\textrm{s}^{-1}$ given in Fig. 1a of~\cite{Boukany2009n2} and $\tau _R^w = 4.1\,\textrm{s}$ quoted in Table 2 of~\cite{Boukany2009n2} (for the sample SBR 250K) gives $\langle\dot{\gamma}\rangle\tau _R^w\simeq 2.9$. The sample SBR 250K (see Table 2 of~\cite{Boukany2009n2}) has $M_w = 250000\,\textrm{g/mol}$ and $M_e = 3300\,\textrm{g/mol}$, which gives $Z = 76$. Then using $Z = 76$, $\langle\dot{\gamma}\rangle\tau _d =200$ and $\tau _d/\tau _R = 3Z$ gives $\langle\dot{\gamma}\rangle\tau _R \simeq 0.95$, which is comparable to the value of $\langle\dot{\gamma}\rangle\tau _R \simeq 1$ specified in case I of the manuscript, this is consistent with Fig. 1 of~\cite{Boukany2009n2}.

\textbf{\textit{Case II: High Shear Rate, Low Strain--}} Similarly, $\langle\dot{\gamma}\rangle = 14\,\textrm{s}^{-1}$ from~\cite{Boukany2009n2} gives $\langle\dot{\gamma}\rangle\tau _R^w \simeq 57$, which is consistent with $\langle\dot{\gamma}\rangle\tau _R > 1$ given in case II of the manuscript and it agrees with Fig. 2 of~\cite{Boukany2009n2}.

\textbf{\textit{Case III: Low Shear Rate, Low Strain--}} Again, $\langle\dot{\gamma}\rangle = 0.05\,\textrm{s}^{-1}$ gives $\langle\dot{\gamma}\rangle\tau _R^w \simeq 0.2$, which is consistent with $\langle\dot{\gamma}\rangle\tau _R < 1$ given in case III of the manuscript, this has close agreement with Fig. 7 of~\cite{Boukany2009n2}.

The shear stresses at the time of shear cessation for the three cases I, II and III are indicated in Fig.~\ref{fig6s}. In case I, the shear stress had gone through the overshoot and it is beginning to decrease. In case two, the flow is switched off before the shear stress reaches the overshoot. Finally, in case III the flow is switched off just before the shear stress reaches the overshoot.
Figure 1(c) of the manuscript shows a comparison of velocity profiles from the simulations and experimental data; the experimental data were obtained from $V_{max}$ in Fig. 1c of~\cite{Boukany2009n2}, made dimensionless using $\hat{V}_{max} = V_{max}\tau/L$, where $\tau = 310\,\textrm{s}$ (from Table 2 of~\cite{Boukany2009n2}) and $L = 0.7\,\textrm{mm}$ as given in section II of~\cite{Boukany2009n2}.
 \begin{figure}[htb]
\begin{center}
{\includegraphics[width=14.0truecm]{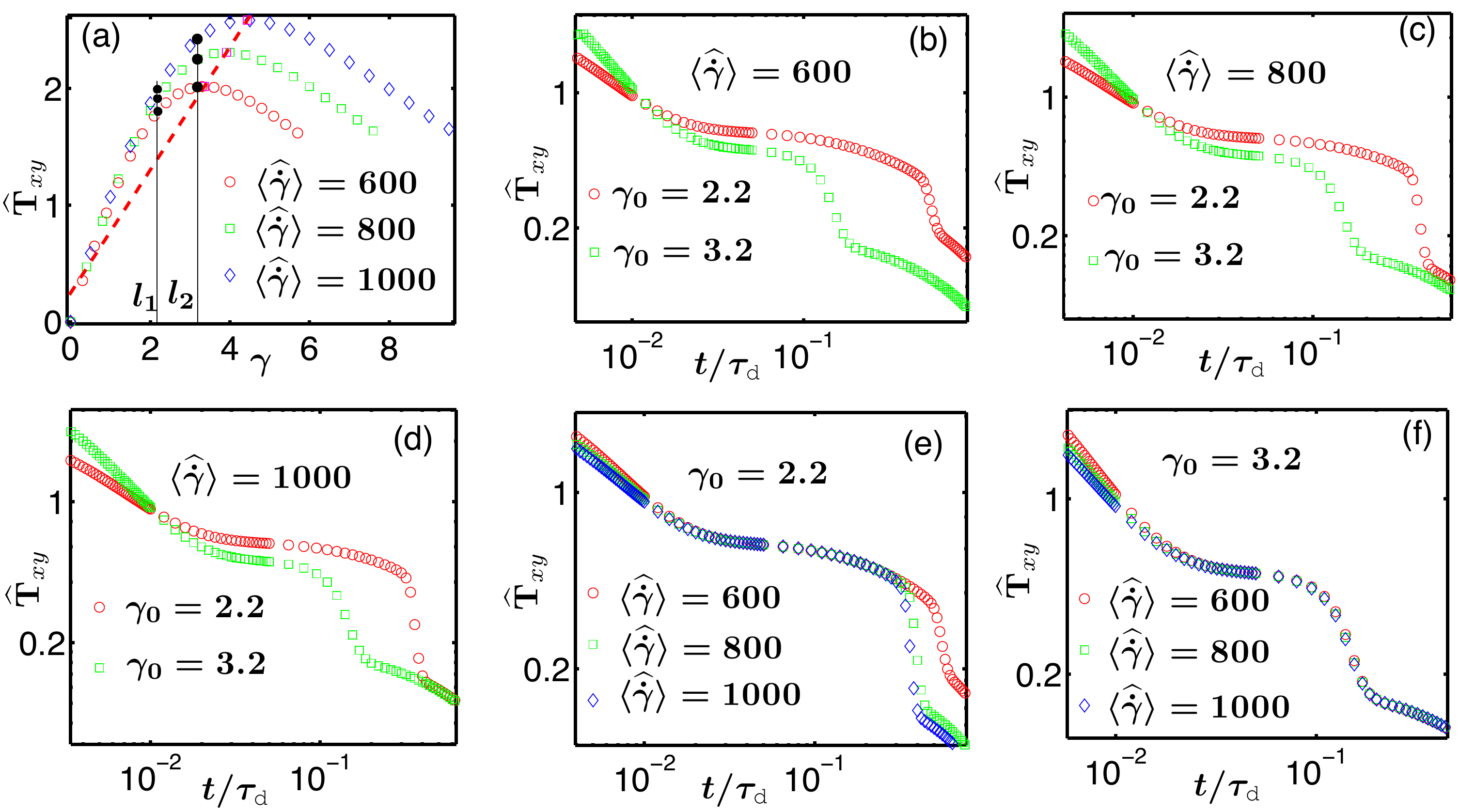}}
\end{center}
\caption{(a)\,Shear stress versus strain at the three different applied shear rates indicated in the figure such that $\langle{\dot{\gamma}}\rangle\tau _R > 1$ in all cases, red circles:\,$\langle{\dot{\gamma}}\rangle\tau _R = 2.8$, green squares:\,$\langle{\dot{\gamma}}\rangle\tau _R = 3.7$ and blue diamonds:\,$\langle\hat{\dot{\gamma}}\rangle\tau _R = 4.6$. The dashed line connects the strains for overshoot and their corresponding stresses for each applied shear rate, while the lines $l_1$ and $l_2$ indicate the applied strains $\gamma _0 = 2.2$ and $\gamma _0 = 3.2$ respectively. (b)-(f)\,Stress relaxation after step strains at different applied strains $\gamma _0$ and shear rates $\langle\hat{\dot{\gamma}}\rangle$ indicated. 
Parameters as in Fig.~\ref{fig4s}.}
 \label{fig7s}
 \end{figure}
 
 \begin{figure}[htb]
\begin{center}
{\includegraphics[width=10.0truecm]{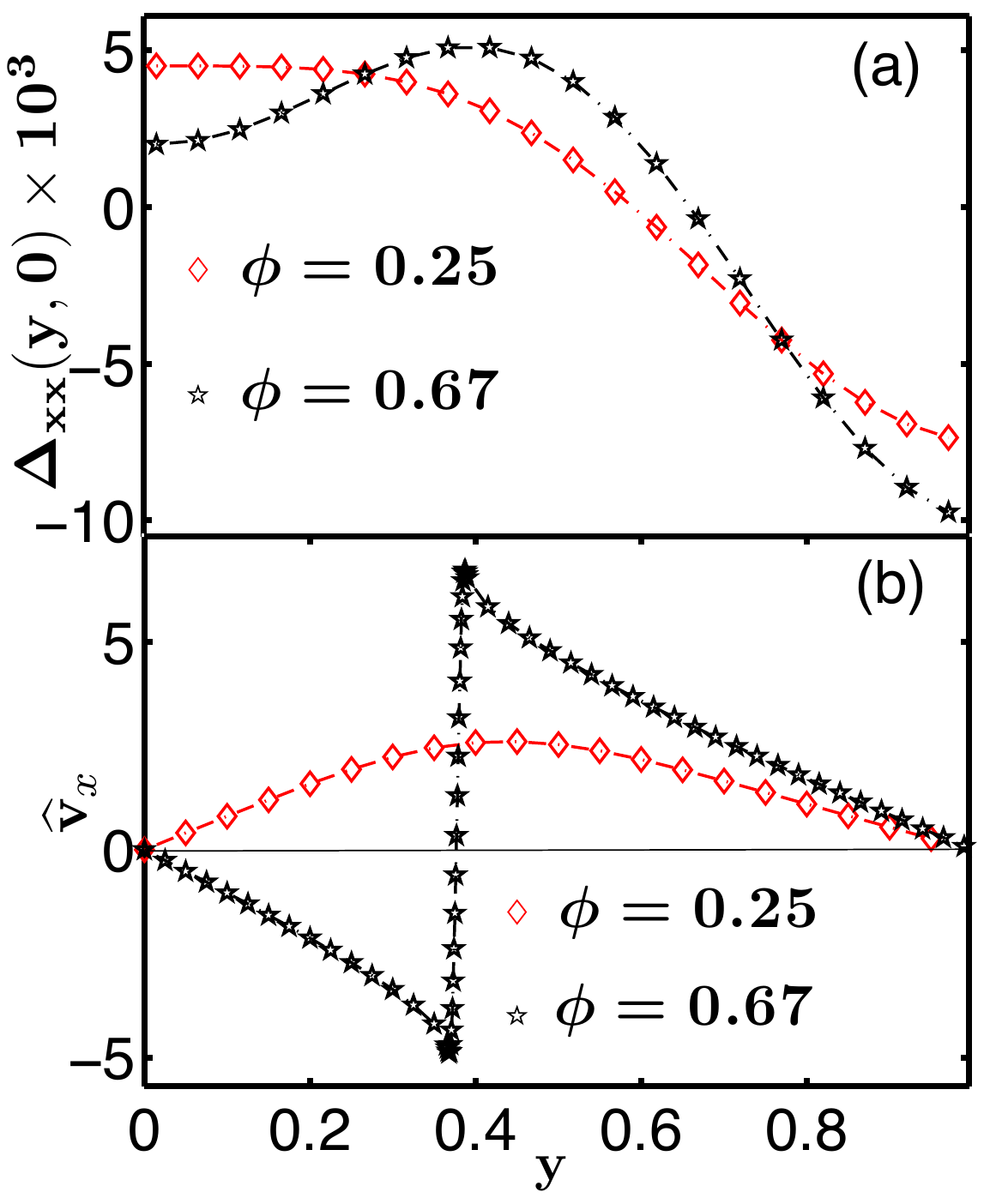}}
\end{center}
\caption{(a)\,Initial perturbation for: $\phi = 0.25$ and $\phi = 0.67$, as shown in the movies \texttt{Recoil.avi} and \texttt{Fracture.avi}. (b)\,Recoil after shear cessation for $\phi = 0.25$ and fracture after shear cessation for $\phi = 0.67$.
Parameters: $\beta = 0$, $Z = \tau _d/(3\tau _R) = 72$, $\langle\hat{\dot{\gamma}}\rangle = 200$, and $\gamma _0 = 2.5$.}
 \label{fig8s}
 \end{figure}

\textbf{\textit{Induction time--}}To check the variation of the delay time after shear cessation before fracture sets in, we performed calculations at three different shear rates satisfying $\langle{\dot{\gamma}}\rangle\tau _R > 1$ (with $\tau _R$ fixed), similar to Fig. 6 of~\cite{Boukany2009n2}. For $\langle\hat{\dot{\gamma}}\rangle = 600$, $\langle{\dot{\gamma}}\rangle\tau _R\simeq 2.8$, $\langle\hat{\dot{\gamma}}\rangle = 800$, $\langle{\dot{\gamma}}\rangle\tau _R\simeq 3.7$ and $\langle\hat{\dot{\gamma}}\rangle = 1000$, $\langle{\dot{\gamma}}\rangle\tau _R\simeq 4.6$. In all cases, the applied strains indicated by the lines $l_1$ and $l_2$ in Fig.~\ref{fig7s}(a), are below the strain for overshoot at the applied shear rate. The overshoot stress is a linear function of the overshoot strain, as  in Fig. 6(a) of~\cite{Boukany2009n2}. Figures~\ref{fig7s}(bcd) show that, for varying strain and given shear rate, the higher plateau stress after stretch relaxation leads to a longer induction time. This characteristic is similar to the the situation in the inset of Fig. 6(b) of~\cite{Boukany2009n2}. 

Figures.~\ref{fig7s}(e-f) show that for fixed strain and varying shear rate, the plateau stresses collapse, and the lower applied shear rate leads to a slightly longer induction time for $\gamma _0 = 2.2$. This can be linked to the faster growth rate $\omega_{\textrm{max}}$ observed for the very high shear rates, in which the viscous contribution to the instability dominates. However, this behaviour does \textit{not} match that displayed in the inset of Fig. 6(b) of~\cite{Boukany2009n2}, in which the \textit{higher} applied shear rate resulted in a larger induction time. We do not have an adequate explanation for these discrepancies. 

\section{Movies}
The movies in \texttt{https://eudoxus.leeds.ac.uk/dynacop/FracturePage.html} illustrate the cases where the fluid undergoes fracture after shear cessation (\texttt{Fracture.avi}) and recoil without fracture (\texttt{Recoil.avi}) for case I. To achieve this, an initial condition of the form $\Delta _{xx}(0,y) = A(\cos (\pi y) + \phi\cos (2\pi y))$ is used to perturb the system. The shape and amplitude of this perturbation can be tuned to bring it close to one of the random perturbations which yields fracture-like behaviour when the component $\Delta _{xx}$ is perturbed. The amplitude is fixed at $A = 0.006$ while the parameter $\phi$ is varied to change the shape of the perturbation. The shapes of this perturbation for $\phi = 0.25$ and $\phi = 0.67$ are shown in Fig.~\ref{fig8s}(a).

For $\phi = 0.67$, the fluid fractures after shear cessation, the window on the left of \texttt{Fracture.avi} shows the fluid velocity from startup (with the upper plate fixed and the lower plate moving) to shear cessation and continues until the end of fracture. Before shear cessation, the fluid is seen to be moving to the left, after which the flow is switched off and the velocity vectors go to zero momentarily (except with a slight bulge due to the initial perturbation). The sizes of the velocity vectors before shear cessation are larger than their sizes after shear cessation by roughly one order of magnitude, hence to make the figure visible in the video, a rescaling of the figure window was carried out after shear cessation. The velocity profile $v$ in the video on the left was made dimensionless using $\hat{v} = v\tau _d/L$. Then using $\tau _d = 310\,\textrm{s}$ and $L = 0.7\,\textrm{mm}$ (from~\cite{Boukany2009n2}) gives the maximum size of velocity vectors $v_{max}$ before shear cessation roughly equal to 0.45\,$\textrm{mm\,s}^{-1}$ and the maximum size after shear cessation is roughly equal to 0.02\,$\textrm{mm\,s}^{-1}$. The velocity profile during fracture is shown in Fig.~\ref{fig8s}(b).

The figure window on the right of \texttt{Fracture.avi} shows the corresponding total shear stress $T_{xy}/G$ from startup until the end of fracture. The total shear stress builds up quickly when the flow is switched on, and then just after the overshoot when the flow is switched off, the total shear stress goes through an initial quick relaxation during which the polymer chains relax stretch. It then enters a slow relaxation when reptation sets in. Although some reptation had already occurred during stretch relaxation, it becomes the dominant mechanism for stress relaxation after stretch relaxation. However, before reptation can completely relax the stress, the growing perturbation causes a sudden quick relaxation of stress. By this time the `fracture plane' is fully developed and the fluid can be seen moving rapidly in two different directions on both sides of this plane. Finally when this rapid motion ceases, the stress resumes its slow relaxation and the material appears to have healed itself.

The case of $\phi = 0.25$, where there is no peak in the initial perturbation as in Fig.~\ref{fig8s}(a), gives a completely different relaxation behaviour in the fluid as shown in \texttt{Recoil.avi}. The left window of that figure shows the fluid velocity from startup to shear cessation and beyond. Like in the case of $\phi = 0.67$, the top plate is fixed while the lower plate moves to the left. After shear cessation, the perturbation is seen to grow for a while but the fluid does not `break' in two unlike in the case of $\phi = 0.67$. The growing perturbation loses the competition against the background reptation and hence the material heals itself and the fluid velocity vanishes after some time. Like in the case of $\phi = 0.67$, the figure window has been rescaled after shear cessation to make the velocity vectors visible. The maximum size of the velocity vectors before shear cessation is roughly equal to 0.45\,$\textrm{mm\,s}^{-1}$ while the maximum size after shear cessation is roughly equal to 0.006\,$\textrm{mm\,s}^{-1}$. The recoil velocity for this case is shown in Fig.~\ref{fig8s}(b).

The right window of \texttt{Recoil.avi} shows the corresponding time dependent total shear stress for this case. It grows quickly from startup like the case of $\phi = 0.67$, then decays quickly during stretch relaxation and ends up with a slow relaxation due to reptation. The stress does not show any stage of rapid relaxation again since reptation is the dominant mechanism for stress relaxation in this case.

The movies were made with a mesh of 100 grid points to reduce the computational time. The relevant parameters were $Z = 72$, $\widehat{\mathcal{D}} = 10^{-5}$, $\epsilon = 10^{-4}$, $\beta = 0$, $\gamma _0 = 2.5$ and $\langle\hat{\dot{\gamma}}\rangle = 200$, which represent case I described in the manuscript.

\renewcommand{\bibname}{References} 
\bibliography{references,mastershear,articles,books} 

\begin{thebibliography}{47}%
\makeatletter
\providecommand \@ifxundefined [1]{%
 \@ifx{#1\undefined}
}%
\providecommand \@ifnum [1]{%
 \ifnum #1\expandafter \@firstoftwo
 \else \expandafter \@secondoftwo
 \fi
}%
\providecommand \@ifx [1]{%
 \ifx #1\expandafter \@firstoftwo
 \else \expandafter \@secondoftwo
 \fi
}%
\providecommand \natexlab [1]{#1}%
\providecommand \enquote  [1]{``#1''}%
\providecommand \bibnamefont  [1]{#1}%
\providecommand \bibfnamefont [1]{#1}%
\providecommand \citenamefont [1]{#1}%
\providecommand \href@noop [0]{\@secondoftwo}%
\providecommand \href [0]{\begingroup \@sanitize@url \@href}%
\providecommand \@href[1]{\@@startlink{#1}\@@href}%
\providecommand \@@href[1]{\endgroup#1\@@endlink}%
\providecommand \@sanitize@url [0]{\catcode `\\12\catcode `\$12\catcode
  `\&12\catcode `\#12\catcode `\^12\catcode `\_12\catcode `\%12\relax}%
\providecommand \@@startlink[1]{}%
\providecommand \@@endlink[0]{}%
\providecommand \url  [0]{\begingroup\@sanitize@url \@url }%
\providecommand \@url [1]{\endgroup\@href {#1}{\urlprefix }}%
\providecommand \urlprefix  [0]{URL }%
\providecommand \Eprint [0]{\href }%
\providecommand \doibase [0]{http://dx.doi.org/}%
\providecommand \selectlanguage [0]{\@gobble}%
\providecommand \bibinfo  [0]{\@secondoftwo}%
\providecommand \bibfield  [0]{\@secondoftwo}%
\providecommand \translation [1]{[#1]}%
\providecommand \BibitemOpen [0]{}%
\providecommand \bibitemStop [0]{}%
\providecommand \bibitemNoStop [0]{.\EOS\space}%
\providecommand \EOS [0]{\spacefactor3000\relax}%
\providecommand \BibitemShut  [1]{\csname bibitem#1\endcsname}%
\let\auto@bib@innerbib\@empty
\bibitem [{\citenamefont {Doyle}\ \emph {et~al.}(1972)\citenamefont {Doyle},
  \citenamefont {Maranci}, \citenamefont {Orowan},\ and\ \citenamefont
  {Stork}}]{Doyle1972}%
  \BibitemOpen
  \bibfield  {author} {\bibinfo {author} {\bibfnamefont {M.~J.}\ \bibnamefont
  {Doyle}}, \bibinfo {author} {\bibfnamefont {A.}~\bibnamefont {Maranci}},
  \bibinfo {author} {\bibfnamefont {E.}~\bibnamefont {Orowan}}, \ and\ \bibinfo
  {author} {\bibfnamefont {S.~T.}\ \bibnamefont {Stork}},\ }\href@noop {}
  {\bibfield  {journal} {\bibinfo  {journal} {Proc. R. Soc. Lond. A}\ }\textbf
  {\bibinfo {volume} {329}},\ \bibinfo {pages} {137} (\bibinfo {year}
  {1972})}\BibitemShut {NoStop}%
\bibitem [{\citenamefont {Lu}\ \emph {et~al.}(2009)\citenamefont {Lu},
  \citenamefont {Ravichandran},\ and\ \citenamefont
  {Johnson}}]{metallicglasses}%
  \BibitemOpen
  \bibfield  {author} {\bibinfo {author} {\bibfnamefont {J.}~\bibnamefont
  {Lu}}, \bibinfo {author} {\bibfnamefont {G.}~\bibnamefont {Ravichandran}}, \
  and\ \bibinfo {author} {\bibfnamefont {W.}~\bibnamefont {Johnson}},\
  }\href@noop {} {\bibfield  {journal} {\bibinfo  {journal} {Acta Mater}\
  }\textbf {\bibinfo {volume} {51}},\ \bibinfo {pages} {3429} (\bibinfo {year}
  {2009})}\BibitemShut {NoStop}%
\bibitem [{\citenamefont {Manning}\ \emph {et~al.}(2007)\citenamefont
  {Manning}, \citenamefont {Langer},\ and\ \citenamefont
  {Carlson}}]{ManLan2007PSNSMP}%
  \BibitemOpen
  \bibfield  {author} {\bibinfo {author} {\bibfnamefont {M.~L.}\ \bibnamefont
  {Manning}}, \bibinfo {author} {\bibfnamefont {J.~S.}\ \bibnamefont {Langer}},
  \ and\ \bibinfo {author} {\bibfnamefont {J.~M.}\ \bibnamefont {Carlson}},\
  }\href@noop {} {\bibfield  {journal} {\bibinfo  {journal} {Phys. Rev. E}\
  }\textbf {\bibinfo {volume} {76}},\ \bibinfo {pages} {056106} (\bibinfo
  {year} {2007})}\BibitemShut {NoStop}%
\bibitem [{\citenamefont {Manning}\ \emph {et~al.}(2009)\citenamefont
  {Manning}, \citenamefont {Daub}, \citenamefont {Langer},\ and\ \citenamefont
  {Carlson}}]{Manning2009rate}%
  \BibitemOpen
  \bibfield  {author} {\bibinfo {author} {\bibfnamefont {M.~L.}\ \bibnamefont
  {Manning}}, \bibinfo {author} {\bibfnamefont {E.~G.}\ \bibnamefont {Daub}},
  \bibinfo {author} {\bibfnamefont {J.~S.}\ \bibnamefont {Langer}}, \ and\
  \bibinfo {author} {\bibfnamefont {J.~M.}\ \bibnamefont {Carlson}},\
  }\href@noop {} {\bibfield  {journal} {\bibinfo  {journal} {Phys. Rev. E}\
  }\textbf {\bibinfo {volume} {79}},\ \bibinfo {pages} {016110} (\bibinfo
  {year} {2009})}\BibitemShut {NoStop}%
\bibitem [{\citenamefont {Furukawa}\ and\ \citenamefont
  {Tanaka}(2009)}]{furukawa2009inhomogeneous}%
  \BibitemOpen
  \bibfield  {author} {\bibinfo {author} {\bibfnamefont {A.}~\bibnamefont
  {Furukawa}}\ and\ \bibinfo {author} {\bibfnamefont {H.}~\bibnamefont
  {Tanaka}},\ }\href@noop {} {\bibfield  {journal} {\bibinfo  {journal} {Nature
  Mat.}\ }\textbf {\bibinfo {volume} {8}},\ \bibinfo {pages} {601} (\bibinfo
  {year} {2009})}\BibitemShut {NoStop}%
\bibitem [{\citenamefont {Boukany}\ \emph {et~al.}(2009)\citenamefont
  {Boukany}, \citenamefont {Wang},\ and\ \citenamefont {Wang}}]{Boukany2009n2}%
  \BibitemOpen
  \bibfield  {author} {\bibinfo {author} {\bibfnamefont {P.~E.}\ \bibnamefont
  {Boukany}}, \bibinfo {author} {\bibfnamefont {S.-Q.}\ \bibnamefont {Wang}}, \
  and\ \bibinfo {author} {\bibfnamefont {X.}~\bibnamefont {Wang}},\ }\href@noop
  {} {\bibfield  {journal} {\bibinfo  {journal} {Macromolecules}\ }\textbf
  {\bibinfo {volume} {42}},\ \bibinfo {pages} {6261} (\bibinfo {year}
  {2009})}\BibitemShut {NoStop}%
\bibitem [{\citenamefont {Fang}\ \emph {et~al.}(2011)\citenamefont {Fang},
  \citenamefont {Wang}, \citenamefont {Tian}, \citenamefont {Wang},
  \citenamefont {Zhu}, \citenamefont {Lin}, \citenamefont {Ma},\ and\
  \citenamefont {Li}}]{FangJOR2011}%
  \BibitemOpen
  \bibfield  {author} {\bibinfo {author} {\bibfnamefont {Y.}~\bibnamefont
  {Fang}}, \bibinfo {author} {\bibfnamefont {G.}~\bibnamefont {Wang}}, \bibinfo
  {author} {\bibfnamefont {N.}~\bibnamefont {Tian}}, \bibinfo {author}
  {\bibfnamefont {X.}~\bibnamefont {Wang}}, \bibinfo {author} {\bibfnamefont
  {X.}~\bibnamefont {Zhu}}, \bibinfo {author} {\bibfnamefont {P.}~\bibnamefont
  {Lin}}, \bibinfo {author} {\bibfnamefont {G.}~\bibnamefont {Ma}}, \ and\
  \bibinfo {author} {\bibfnamefont {L.}~\bibnamefont {Li}},\ }\href@noop {}
  {\bibfield  {journal} {\bibinfo  {journal} {J. Rheol.}\ }\textbf {\bibinfo
  {volume} {55}},\ \bibinfo {pages} {939} (\bibinfo {year} {2011})}\BibitemShut
  {NoStop}%
\bibitem [{\citenamefont {Doi}\ and\ \citenamefont
  {Edwards}(1989)}]{doiedwards}%
  \BibitemOpen
  \bibfield  {author} {\bibinfo {author} {\bibfnamefont {M.}~\bibnamefont
  {Doi}}\ and\ \bibinfo {author} {\bibfnamefont {S.~F.}\ \bibnamefont
  {Edwards}},\ }\href@noop {} {\emph {\bibinfo {title} {The Theory of Polymer
  Dynamics}}}\ (\bibinfo  {publisher} {Oxford University Press},\ \bibinfo
  {address} {Oxford},\ \bibinfo {year} {1989})\BibitemShut {NoStop}%
\bibitem [{\citenamefont {Marrucci}\ and\ \citenamefont
  {Grizzuti}(1983)}]{Marrucci1983}%
  \BibitemOpen
  \bibfield  {author} {\bibinfo {author} {\bibfnamefont {G.}~\bibnamefont
  {Marrucci}}\ and\ \bibinfo {author} {\bibfnamefont {N.}~\bibnamefont
  {Grizzuti}},\ }\href@noop {} {\bibfield  {journal} {\bibinfo  {journal} {J.
  Rheol.}\ }\textbf {\bibinfo {volume} {27}},\ \bibinfo {pages} {433} (\bibinfo
  {year} {1983})}\BibitemShut {NoStop}%
\bibitem [{\citenamefont {Morrison}\ and\ \citenamefont
  {Larson}(1992)}]{Morrison1992}%
  \BibitemOpen
  \bibfield  {author} {\bibinfo {author} {\bibfnamefont {F.~A.}\ \bibnamefont
  {Morrison}}\ and\ \bibinfo {author} {\bibfnamefont {R.~G.}\ \bibnamefont
  {Larson}},\ }\href@noop {} {\bibfield  {journal} {\bibinfo  {journal} {J.
  Polym. Sci.: B}\ }\textbf {\bibinfo {volume} {30}},\ \bibinfo {pages} {943}
  (\bibinfo {year} {1992})}\BibitemShut {NoStop}%
\bibitem [{\citenamefont {Adams}\ and\ \citenamefont
  {Olmsted}(2009{\natexlab{a}})}]{Adams2009}%
  \BibitemOpen
  \bibfield  {author} {\bibinfo {author} {\bibfnamefont {J.~M.}\ \bibnamefont
  {Adams}}\ and\ \bibinfo {author} {\bibfnamefont {P.~D.}\ \bibnamefont
  {Olmsted}},\ }\href@noop {} {\bibfield  {journal} {\bibinfo  {journal} {Phys.
  Rev. Lett.}\ }\textbf {\bibinfo {volume} {102}},\ \bibinfo {pages} {067801}
  (\bibinfo {year} {2009}{\natexlab{a}})}\BibitemShut {NoStop}%
\bibitem [{\citenamefont {Cao}\ and\ \citenamefont
  {Likhtman}(2012)}]{CaoLikhtmanPRL2012}%
  \BibitemOpen
  \bibfield  {author} {\bibinfo {author} {\bibfnamefont {J.}~\bibnamefont
  {Cao}}\ and\ \bibinfo {author} {\bibfnamefont {A.~E.}\ \bibnamefont
  {Likhtman}},\ }\href {\doibase 10.1103/PhysRevLett.108.028302} {\bibfield
  {journal} {\bibinfo  {journal} {Phys. Rev. Lett.}\ }\textbf {\bibinfo
  {volume} {108}},\ \bibinfo {pages} {028302} (\bibinfo {year}
  {2012})}\BibitemShut {NoStop}%
\bibitem [{\citenamefont {Menezes}\ and\ \citenamefont
  {Graessley}(1982)}]{Menezes1982}%
  \BibitemOpen
  \bibfield  {author} {\bibinfo {author} {\bibfnamefont {E.~V.}\ \bibnamefont
  {Menezes}}\ and\ \bibinfo {author} {\bibfnamefont {W.~W.}\ \bibnamefont
  {Graessley}},\ }\href@noop {} {\bibfield  {journal} {\bibinfo  {journal} {J.
  Polym. Sci.: B}\ }\textbf {\bibinfo {volume} {20}},\ \bibinfo {pages} {1817}
  (\bibinfo {year} {1982})}\BibitemShut {NoStop}%
\bibitem [{\citenamefont {Spenley}\ \emph {et~al.}(1993)\citenamefont
  {Spenley}, \citenamefont {Cates},\ and\ \citenamefont
  {McLeish}}]{Spenley1993}%
  \BibitemOpen
  \bibfield  {author} {\bibinfo {author} {\bibfnamefont {N.~A.}\ \bibnamefont
  {Spenley}}, \bibinfo {author} {\bibfnamefont {M.~E.}\ \bibnamefont {Cates}},
  \ and\ \bibinfo {author} {\bibfnamefont {T.~C.~B.}\ \bibnamefont {McLeish}},\
  }\href@noop {} {\bibfield  {journal} {\bibinfo  {journal} {Phys. Rev. Lett.}\
  }\textbf {\bibinfo {volume} {71}},\ \bibinfo {pages} {939} (\bibinfo {year}
  {1993})}\BibitemShut {NoStop}%
\bibitem [{\citenamefont {Olmsted}(2008)}]{Olmsted2008}%
  \BibitemOpen
  \bibfield  {author} {\bibinfo {author} {\bibfnamefont {P.~D.}\ \bibnamefont
  {Olmsted}},\ }\href@noop {} {\bibfield  {journal} {\bibinfo  {journal}
  {Rheol. Acta}\ }\textbf {\bibinfo {volume} {47}},\ \bibinfo {pages} {283}
  (\bibinfo {year} {2008})}\BibitemShut {NoStop}%
\bibitem [{\citenamefont {Lu}\ \emph {et~al.}(2000)\citenamefont {Lu},
  \citenamefont {Olmsted},\ and\ \citenamefont {Ball}}]{Lu2000}%
  \BibitemOpen
  \bibfield  {author} {\bibinfo {author} {\bibfnamefont {C.-Y.~D.}\
  \bibnamefont {Lu}}, \bibinfo {author} {\bibfnamefont {P.~D.}\ \bibnamefont
  {Olmsted}}, \ and\ \bibinfo {author} {\bibfnamefont {R.~C.}\ \bibnamefont
  {Ball}},\ }\href@noop {} {\bibfield  {journal} {\bibinfo  {journal} {Phys.
  Rev. Lett.}\ }\textbf {\bibinfo {volume} {84}},\ \bibinfo {pages} {642}
  (\bibinfo {year} {2000})}\BibitemShut {NoStop}%
\bibitem [{\citenamefont {Huseby}(1966)}]{huseby1966hypothesis}%
  \BibitemOpen
  \bibfield  {author} {\bibinfo {author} {\bibfnamefont {T.~W.}\ \bibnamefont
  {Huseby}},\ }\href@noop {} {\bibfield  {journal} {\bibinfo  {journal} {J.
  Rheol.}\ }\textbf {\bibinfo {volume} {10}},\ \bibinfo {pages} {181} (\bibinfo
  {year} {1966})}\BibitemShut {NoStop}%
\bibitem [{\citenamefont {Lin}(1985)}]{lin1985explanation}%
  \BibitemOpen
  \bibfield  {author} {\bibinfo {author} {\bibfnamefont {Y.~H.}\ \bibnamefont
  {Lin}},\ }\href@noop {} {\bibfield  {journal} {\bibinfo  {journal} {J.
  Rheol.}\ }\textbf {\bibinfo {volume} {29}},\ \bibinfo {pages} {605} (\bibinfo
  {year} {1985})}\BibitemShut {NoStop}%
\bibitem [{\citenamefont {McLeish}\ and\ \citenamefont
  {Ball}(1986)}]{mcleish86}%
  \BibitemOpen
  \bibfield  {author} {\bibinfo {author} {\bibfnamefont {T.~C.~B.}\
  \bibnamefont {McLeish}}\ and\ \bibinfo {author} {\bibfnamefont {R.~C.}\
  \bibnamefont {Ball}},\ }\href@noop {} {\bibfield  {journal} {\bibinfo
  {journal} {J.~Poly. Sci. B-Poly. Phys.}\ }\textbf {\bibinfo {volume} {24}},\
  \bibinfo {pages} {1735} (\bibinfo {year} {1986})}\BibitemShut {NoStop}%
\bibitem [{\citenamefont {McLeish}(1987)}]{mcleish87}%
  \BibitemOpen
  \bibfield  {author} {\bibinfo {author} {\bibfnamefont {T.~C.~B.}\
  \bibnamefont {McLeish}},\ }\href@noop {} {\bibfield  {journal} {\bibinfo
  {journal} {J.~Poly. Sci. B-Poly. Phys.}\ }\textbf {\bibinfo {volume} {25}},\
  \bibinfo {pages} {2253} (\bibinfo {year} {1987})}\BibitemShut {NoStop}%
\bibitem [{\citenamefont {Malkus}\ \emph {et~al.}(1991)\citenamefont {Malkus},
  \citenamefont {Nohel},\ and\ \citenamefont {Plohr}}]{MNP91}%
  \BibitemOpen
  \bibfield  {author} {\bibinfo {author} {\bibfnamefont {D.~S.}\ \bibnamefont
  {Malkus}}, \bibinfo {author} {\bibfnamefont {J.~A.}\ \bibnamefont {Nohel}}, \
  and\ \bibinfo {author} {\bibfnamefont {B.~J.}\ \bibnamefont {Plohr}},\
  }\href@noop {} {\bibfield  {journal} {\bibinfo  {journal} {Siam J. Appl.
  Math.}\ }\textbf {\bibinfo {volume} {51}},\ \bibinfo {pages} {899} (\bibinfo
  {year} {1991})}\BibitemShut {NoStop}%
\bibitem [{\citenamefont {Denn}(1990)}]{Denn90}%
  \BibitemOpen
  \bibfield  {author} {\bibinfo {author} {\bibfnamefont {M.~M.}\ \bibnamefont
  {Denn}},\ }\href@noop {} {\bibfield  {journal} {\bibinfo  {journal} {Ann.
  Rev. Fluid Mech.}\ }\textbf {\bibinfo {volume} {22}},\ \bibinfo {pages} {13}
  (\bibinfo {year} {1990})}\BibitemShut {NoStop}%
\bibitem [{\citenamefont {Vinogradov}\ \emph {et~al.}(1972)\citenamefont
  {Vinogradov}, \citenamefont {Malkin}, \citenamefont {Yanovskii},
  \citenamefont {Borisenkova}, \citenamefont {Yarlykov},\ and\ \citenamefont
  {Berezhnaya}}]{Vinogradov1972}%
  \BibitemOpen
  \bibfield  {author} {\bibinfo {author} {\bibfnamefont {G.~V.}\ \bibnamefont
  {Vinogradov}}, \bibinfo {author} {\bibfnamefont {A.~Y.}\ \bibnamefont
  {Malkin}}, \bibinfo {author} {\bibfnamefont {Y.~G.}\ \bibnamefont
  {Yanovskii}}, \bibinfo {author} {\bibfnamefont {E.~K.}\ \bibnamefont
  {Borisenkova}}, \bibinfo {author} {\bibfnamefont {B.~V.}\ \bibnamefont
  {Yarlykov}}, \ and\ \bibinfo {author} {\bibfnamefont {G.~V.}\ \bibnamefont
  {Berezhnaya}},\ }\href@noop {} {\bibfield  {journal} {\bibinfo  {journal} {J.
  Polym. Sci.: Part A-2}\ }\textbf {\bibinfo {volume} {10}},\ \bibinfo {pages}
  {1061} (\bibinfo {year} {1972})}\BibitemShut {NoStop}%
\bibitem [{\citenamefont {Lim}\ and\ \citenamefont
  {Schowalter}(1989)}]{lim1989wall}%
  \BibitemOpen
  \bibfield  {author} {\bibinfo {author} {\bibfnamefont {F.~J.}\ \bibnamefont
  {Lim}}\ and\ \bibinfo {author} {\bibfnamefont {W.~R.}\ \bibnamefont
  {Schowalter}},\ }\href@noop {} {\bibfield  {journal} {\bibinfo  {journal} {J.
  Rheology}\ }\textbf {\bibinfo {volume} {33}},\ \bibinfo {pages} {1359}
  (\bibinfo {year} {1989})}\BibitemShut {NoStop}%
\bibitem [{\citenamefont {Wang}(1999)}]{Wan1999PCE}%
  \BibitemOpen
  \bibfield  {author} {\bibinfo {author} {\bibfnamefont {S.~Q.}\ \bibnamefont
  {Wang}},\ }in\ \href@noop {} {\emph {\bibinfo {booktitle} {Polymers in
  Confined Environments}}},\ \bibinfo {series} {Adv. Poly. Sci.}, Vol.\
  \bibinfo {volume} {138}\ (\bibinfo  {publisher} {Springer},\ \bibinfo
  {address} {Berlin},\ \bibinfo {year} {1999})\ pp.\ \bibinfo {pages}
  {227--275}\BibitemShut {NoStop}%
\bibitem [{\citenamefont {Denn}(2001)}]{Den2001ARFM}%
  \BibitemOpen
  \bibfield  {author} {\bibinfo {author} {\bibfnamefont {M.~M.}\ \bibnamefont
  {Denn}},\ }\href@noop {} {\bibfield  {journal} {\bibinfo  {journal} {Ann.
  Rev. Fl. Mech.}\ }\textbf {\bibinfo {volume} {33}},\ \bibinfo {pages} {265}
  (\bibinfo {year} {2001})}\BibitemShut {NoStop}%
\bibitem [{\citenamefont {Ianniruberto}\ and\ \citenamefont
  {Marrucci}(1996)}]{Ianniruberto1996}%
  \BibitemOpen
  \bibfield  {author} {\bibinfo {author} {\bibfnamefont {G.}~\bibnamefont
  {Ianniruberto}}\ and\ \bibinfo {author} {\bibfnamefont {G.}~\bibnamefont
  {Marrucci}},\ }\href@noop {} {\bibfield  {journal} {\bibinfo  {journal} {J.
  Non-Newt. Fl. Mech.}\ }\textbf {\bibinfo {volume} {65}},\ \bibinfo {pages}
  {241} (\bibinfo {year} {1996})}\BibitemShut {NoStop}%
\bibitem [{\citenamefont {Mead}\ and\ \citenamefont {Larson}(1998)}]{Mead1998}%
  \BibitemOpen
  \bibfield  {author} {\bibinfo {author} {\bibfnamefont {D.~W.}\ \bibnamefont
  {Mead}}\ and\ \bibinfo {author} {\bibfnamefont {R.~G.}\ \bibnamefont
  {Larson}},\ }\href@noop {} {\bibfield  {journal} {\bibinfo  {journal}
  {Macromolecules}\ }\textbf {\bibinfo {volume} {31}},\ \bibinfo {pages} {7895}
  (\bibinfo {year} {1998})}\BibitemShut {NoStop}%
\bibitem [{\citenamefont {Hu}\ \emph {et~al.}(2007)\citenamefont {Hu},
  \citenamefont {Wilen}, \citenamefont {Philips},\ and\ \citenamefont
  {Lips}}]{hu2007cre}%
  \BibitemOpen
  \bibfield  {author} {\bibinfo {author} {\bibfnamefont {Y.~T.}\ \bibnamefont
  {Hu}}, \bibinfo {author} {\bibfnamefont {L.}~\bibnamefont {Wilen}}, \bibinfo
  {author} {\bibfnamefont {A.}~\bibnamefont {Philips}}, \ and\ \bibinfo
  {author} {\bibfnamefont {A.}~\bibnamefont {Lips}},\ }\href@noop {} {\bibfield
   {journal} {\bibinfo  {journal} {J. Rheol.}\ }\textbf {\bibinfo {volume}
  {51}},\ \bibinfo {pages} {275} (\bibinfo {year} {2007})}\BibitemShut
  {NoStop}%
\bibitem [{\citenamefont {Ravindranath}\ \emph {et~al.}(2008)\citenamefont
  {Ravindranath}, \citenamefont {Wang}, \citenamefont {Olechnowicz},\ and\
  \citenamefont {Quirk}}]{Ravindranath2008}%
  \BibitemOpen
  \bibfield  {author} {\bibinfo {author} {\bibfnamefont {S.}~\bibnamefont
  {Ravindranath}}, \bibinfo {author} {\bibfnamefont {S.-Q.}\ \bibnamefont
  {Wang}}, \bibinfo {author} {\bibfnamefont {M.}~\bibnamefont {Olechnowicz}}, \
  and\ \bibinfo {author} {\bibfnamefont {R.}~\bibnamefont {Quirk}},\
  }\href@noop {} {\bibfield  {journal} {\bibinfo  {journal} {Macromolecules}\
  }\textbf {\bibinfo {volume} {41}},\ \bibinfo {pages} {2663} (\bibinfo {year}
  {2008})}\BibitemShut {NoStop}%
\bibitem [{\citenamefont {Tapadia}\ \emph {et~al.}(2006)\citenamefont
  {Tapadia}, \citenamefont {Ravindranath},\ and\ \citenamefont
  {Wang}}]{Tapadia2006n1}%
  \BibitemOpen
  \bibfield  {author} {\bibinfo {author} {\bibfnamefont {P.}~\bibnamefont
  {Tapadia}}, \bibinfo {author} {\bibfnamefont {S.}~\bibnamefont
  {Ravindranath}}, \ and\ \bibinfo {author} {\bibfnamefont {S.-Q.}\
  \bibnamefont {Wang}},\ }\href@noop {} {\bibfield  {journal} {\bibinfo
  {journal} {Phys. Rev. Lett.}\ }\textbf {\bibinfo {volume} {96}},\ \bibinfo
  {pages} {196001} (\bibinfo {year} {2006})}\BibitemShut {NoStop}%
\bibitem [{\citenamefont {Hu}(2010)}]{Hu2010}%
  \BibitemOpen
  \bibfield  {author} {\bibinfo {author} {\bibfnamefont {Y.~T.}\ \bibnamefont
  {Hu}},\ }\href@noop {} {\bibfield  {journal} {\bibinfo  {journal} {J.
  Rheol.}\ }\textbf {\bibinfo {volume} {54}},\ \bibinfo {pages} {1307}
  (\bibinfo {year} {2010})}\BibitemShut {NoStop}%
\bibitem [{\citenamefont {Adams}\ and\ \citenamefont
  {Olmsted}(2009{\natexlab{b}})}]{Adams2009b}%
  \BibitemOpen
  \bibfield  {author} {\bibinfo {author} {\bibfnamefont {J.~M.}\ \bibnamefont
  {Adams}}\ and\ \bibinfo {author} {\bibfnamefont {P.~D.}\ \bibnamefont
  {Olmsted}},\ }\href@noop {} {\bibfield  {journal} {\bibinfo  {journal} {Phys.
  Rev. Lett.}\ }\textbf {\bibinfo {volume} {103}},\ \bibinfo {pages} {219802}
  (\bibinfo {year} {2009}{\natexlab{b}})}\BibitemShut {NoStop}%
\bibitem [{\citenamefont {Wang}(2009)}]{Wang2009}%
  \BibitemOpen
  \bibfield  {author} {\bibinfo {author} {\bibfnamefont {S.-Q.}\ \bibnamefont
  {Wang}},\ }\href@noop {} {\bibfield  {journal} {\bibinfo  {journal} {Phys.
  Rev. Lett.}\ }\textbf {\bibinfo {volume} {103}},\ \bibinfo {pages} {219801}
  (\bibinfo {year} {2009})}\BibitemShut {NoStop}%
\bibitem [{\citenamefont {Likhtman}\ and\ \citenamefont
  {Graham}(2003)}]{Likhtman2003}%
  \BibitemOpen
  \bibfield  {author} {\bibinfo {author} {\bibfnamefont {A.~E.}\ \bibnamefont
  {Likhtman}}\ and\ \bibinfo {author} {\bibfnamefont {R.~S.}\ \bibnamefont
  {Graham}},\ }\href@noop {} {\bibfield  {journal} {\bibinfo  {journal} {J.
  Non-Newt. Fl. Mech.}\ }\textbf {\bibinfo {volume} {114}},\ \bibinfo {pages}
  {1} (\bibinfo {year} {2003})}\BibitemShut {NoStop}%
\bibitem [{\citenamefont {Graham}\ \emph {et~al.}(2003)\citenamefont {Graham},
  \citenamefont {Likhtman}, \citenamefont {McLeish},\ and\ \citenamefont
  {Milner}}]{Graham2003}%
  \BibitemOpen
  \bibfield  {author} {\bibinfo {author} {\bibfnamefont {R.~S.}\ \bibnamefont
  {Graham}}, \bibinfo {author} {\bibfnamefont {A.~E.}\ \bibnamefont
  {Likhtman}}, \bibinfo {author} {\bibfnamefont {T.~C.~B.}\ \bibnamefont
  {McLeish}}, \ and\ \bibinfo {author} {\bibfnamefont {S.~T.}\ \bibnamefont
  {Milner}},\ }\href@noop {} {\bibfield  {journal} {\bibinfo  {journal} {J.
  Rheol.}\ }\textbf {\bibinfo {volume} {47}},\ \bibinfo {pages} {1171}
  (\bibinfo {year} {2003})}\BibitemShut {NoStop}%
\bibitem [{\citenamefont {Tapadia}\ and\ \citenamefont
  {Wang}(2003)}]{Tapadia2003}%
  \BibitemOpen
  \bibfield  {author} {\bibinfo {author} {\bibfnamefont {P.}~\bibnamefont
  {Tapadia}}\ and\ \bibinfo {author} {\bibfnamefont {S.-Q.}\ \bibnamefont
  {Wang}},\ }\href@noop {} {\bibfield  {journal} {\bibinfo  {journal} {Phys.
  Rev. Lett.}\ }\textbf {\bibinfo {volume} {91}},\ \bibinfo {pages} {198301}
  (\bibinfo {year} {2003})}\BibitemShut {NoStop}%
\bibitem [{\citenamefont {Adams}\ \emph {et~al.}(2008)\citenamefont {Adams},
  \citenamefont {Fielding},\ and\ \citenamefont {Olmsted}}]{Adams2008}%
  \BibitemOpen
  \bibfield  {author} {\bibinfo {author} {\bibfnamefont {J.~M.}\ \bibnamefont
  {Adams}}, \bibinfo {author} {\bibfnamefont {S.~M.}\ \bibnamefont {Fielding}},
  \ and\ \bibinfo {author} {\bibfnamefont {P.~D.}\ \bibnamefont {Olmsted}},\
  }\href@noop {} {\bibfield  {journal} {\bibinfo  {journal} {J. Non-Newt. Fl.
  Mech.}\ }\textbf {\bibinfo {volume} {151}},\ \bibinfo {pages} {101} (\bibinfo
  {year} {2008})}\BibitemShut {NoStop}%
\bibitem [{\citenamefont {Adams}\ \emph {et~al.}(2011)\citenamefont {Adams},
  \citenamefont {Fielding},\ and\ \citenamefont {Olmsted}}]{adams2011}%
  \BibitemOpen
  \bibfield  {author} {\bibinfo {author} {\bibfnamefont {J.~M.}\ \bibnamefont
  {Adams}}, \bibinfo {author} {\bibfnamefont {S.~M.}\ \bibnamefont {Fielding}},
  \ and\ \bibinfo {author} {\bibfnamefont {P.~D.}\ \bibnamefont {Olmsted}},\
  }\href@noop {} {\bibfield  {journal} {\bibinfo  {journal} {J. Rheol.}\
  }\textbf {\bibinfo {volume} {55}},\ \bibinfo {pages} {1007} (\bibinfo {year}
  {2011})}\BibitemShut {NoStop}%
\bibitem [{\citenamefont {Moorcroft}\ and\ \citenamefont
  {Fielding}(2013)}]{moorcroft2013}%
  \BibitemOpen
  \bibfield  {author} {\bibinfo {author} {\bibfnamefont {R.~L.}\ \bibnamefont
  {Moorcroft}}\ and\ \bibinfo {author} {\bibfnamefont {S.~M.}\ \bibnamefont
  {Fielding}},\ }\href@noop {} {\bibfield  {journal} {\bibinfo  {journal}
  {Phys. Rev. Lett.}\ }\textbf {\bibinfo {volume} {110}},\ \bibinfo {pages}
  {086001} (\bibinfo {year} {2013})}\BibitemShut {NoStop}%
\bibitem [{Sup()}]{Supplementary}%
  \BibitemOpen
  \href@noop {} {}\bibinfo {note} {The Supplementary Information contains
  details about the stability calculation, the effects of different initial
  conditions; and "movies" (also available at
  \url{https://eudoxus.leeds.ac.uk/dynacop/FracturePage.html}).}\BibitemShut
  {Stop}%
\bibitem [{Note1()}]{Note1}%
  \BibitemOpen
  \bibinfo {note} {The definition of unrelaxed segments $\mu (t)$ matches the
  linear relaxation function $G(t)\equiv \protect \qopname \relax
  m{lim}_{\gamma _0\rightarrow 0}T_{x,y}(t,\gamma )/\gamma _0$, as does the
  equivalent function used by Marrucci and Grizzuti for the DE model \cite
  {Marrucci1983}}\BibitemShut {NoStop}%
\bibitem [{ene()}]{energyfunction}%
  \BibitemOpen
  \href@noop {} {}\bibinfo {note} {Following \cite{Marrucci1983} we use the
  stored free energy due to the orientational distribution of tube segments,
  given within the independent alignment approximation by
  $F(\gamma)=\tfrac12\int_0^1\ln\left\{\tfrac12
  \left(1+\gamma^2x^2+\left[(x^2\gamma^2-1)^2 +
  4\gamma^2x^4\right]^{1/2}\right)\right\}\,dx$.}\BibitemShut {Stop}%
\bibitem [{\citenamefont {Moorcroft}\ \emph {et~al.}(2011)\citenamefont
  {Moorcroft}, \citenamefont {Cates},\ and\ \citenamefont
  {Fielding}}]{Moorcroft2011}%
  \BibitemOpen
  \bibfield  {author} {\bibinfo {author} {\bibfnamefont {R.~L.}\ \bibnamefont
  {Moorcroft}}, \bibinfo {author} {\bibfnamefont {M.~E.}\ \bibnamefont
  {Cates}}, \ and\ \bibinfo {author} {\bibfnamefont {S.~M.}\ \bibnamefont
  {Fielding}},\ }\href@noop {} {\bibfield  {journal} {\bibinfo  {journal}
  {Phys. Rev. Lett.}\ }\textbf {\bibinfo {volume} {106}},\ \bibinfo {pages}
  {055502} (\bibinfo {year} {2011})}\BibitemShut {NoStop}%
\bibitem [{\citenamefont {Likhtman}(2009)}]{likhtman2009whither}%
  \BibitemOpen
  \bibfield  {author} {\bibinfo {author} {\bibfnamefont {A.}~\bibnamefont
  {Likhtman}},\ }\href@noop {} {\bibfield  {journal} {\bibinfo  {journal} {J
  Non-{N}ewt.Fl. Mech.}\ }\textbf {\bibinfo {volume} {157}},\ \bibinfo {pages}
  {158} (\bibinfo {year} {2009})}\BibitemShut {NoStop}%
\bibitem [{Note2()}]{Note2}%
  \BibitemOpen
  \bibinfo {note} {The time $t$ and spatial variable $y$ have been made
  dimensionless as $\protect \mathaccentV {hat}05E{t} = t/\tau _d$ and
  $\protect \mathaccentV {hat}05E{y} = y/L$ respectively. However, for
  simplicity the quantities $\protect \mathaccentV {hat}05E{t}$ and $\protect
  \mathaccentV {hat}05E{y}$ have been written as $t$ and $y$ in Eq.~\ref
  {eq2}}\BibitemShut {NoStop}%
\bibitem [{\citenamefont {Fielding}\ and\ \citenamefont
  {Olmsted}(2003)}]{fielding03a}%
  \BibitemOpen
  \bibfield  {author} {\bibinfo {author} {\bibfnamefont {S.~M.}\ \bibnamefont
  {Fielding}}\ and\ \bibinfo {author} {\bibfnamefont {P.~D.}\ \bibnamefont
  {Olmsted}},\ }\href@noop {} {\bibfield  {journal} {\bibinfo  {journal} {Phys.
  Rev. Lett.}\ }\textbf {\bibinfo {volume} {90}},\ \bibinfo {pages} {224501}
  (\bibinfo {year} {2003})}\BibitemShut {NoStop}%
\end{thebibliography}%
\end{document}